\documentclass[preprint]{elsarticle}
\usepackage{hyperref}
\usepackage{color}
\usepackage{graphicx}
\usepackage{grffile}
\usepackage{longtable}
\usepackage{wrapfig}
\usepackage{rotating}
\usepackage[normalem]{ulem}
\usepackage{amsmath}
\usepackage{textcomp}
\usepackage{amssymb}
\usepackage{capt-of}
\usepackage{hyperref}
\usepackage{listingsutf8}
\usepackage[T1]{fontenc}
\journal{NIMA}

\begin{document}

\begin{frontmatter}

\title{Performance of The Advanced GAmma Tracking Array at GANIL}

\author[csnsm]{J. Ljungvall
  \corref{mycorrespondingauthor}}
\cortext[mycorrespondingauthor]
        {Corresponding author: joa.ljungvall@csnsm.in2p3.fr}

\author[csic]{R.M. P{\'e}rez-Vidal}
\author[csnsm]{A. Lopez-Martens}
\author[ganil,ill]{C. Michelagnoli}
\author[ganil]{E. Cl{\'e}ment}
\author[csnsm,lyon]{J. Dudouet}
\author[csic]{A. Gadea}
\author[koln]{H. Hess}
\author[csnsm]{A. Korichi}
\author[stfc]{M. Labiche}
\author[lund]{N. Lalovi{\'c}}
\author[ganil]{H. J. Li}
\author[padova1,padova2]{F. Recchia}
\author[]{and the AGATA collaboration}
\address[csnsm]{CSNSM, Universit{\'e} Paris-Sud, CNRS/IN2P3, Universit{\'e}
  Paris-Saclay, 91405 Orsay, France}
\address[csic]{Instituto de F{\'i}sica Corpuscular, CSIC - Universidad de
  Valencia, E-46980 Paterna, Valencia, Spain}
\address[ganil]{GANIL,
  CEA/DRF-CNRS/IN2P3, BP 55027, 14076 Caen cedex 5, France }
\address[ill]{Institut Laue-Langevin, B.P. 156, F-38042
  Grenoble cedex 9, France}
\address[lund]{Department of Physics, Lund University, SE-22100 Lund, Sweden}
\address[koln]{IKP, University of Cologne, D-50937 Cologne, Germany}
\address[padova1]{INFN Sezione di Padova, I-35131 Padova, Italy}
\address[padova2]{Dipartimento di Fisica e Astronomia dell'Universit{\`a} di Padova, I-35131 Padova, Italy }
\address[lyon]{Institut de Physique Nucl{\'e}aire de Lyon, Universit{\'e} de
  Lyon, Universit{\'e} Lyon 1, CNRS-IN2P3, F-69622 Villeurbanne, France}
\address[stfc]{STFC Daresbury Laboratory, Daresbury, Warrington WA4
  4AD, United Kingdom}
\begin{abstract}
  The performance of the Advanced GAmma Tracking Array (AGATA) at
  GANIL is discussed, on the basis of the analysis of source and
  in-beam data taken with up to 30 segmented crystals. Data processing
  is described in detail. The performance of individual detectors are
  shown. The efficiency of the individual detectors as well as the
  efficiency after $\gamma$-ray tracking are discussed. Recent
  developments of $\gamma$-ray tracking are also presented. The
  experimentally achieved peak-to-total is compared with simulations
  showing the impact of back-scattered $\gamma$ rays on the
  peak-to-total in a $\gamma$-ray tracking array. An estimate of the
  achieved position resolution using the Doppler broadening of in-beam
  data is also given.

  Angular correlations from source measurements are shown together
  with different methods to take into account the effects of
  $\gamma$-ray tracking on the normalization of the angular
  correlations.

\end{abstract}


\end{frontmatter}


\section{Introduction}
In order to perform $\gamma$-ray spectroscopy nuclear structure
studies in conditions of extreme neutron/proton asymmetry and/or
extreme angular momentum the so-called $\gamma$-ray tracking arrays
are considered as indispensable tools. Two international
collaborations, Advanced-GAmma-Tracking-Array (AGATA)
\cite{Akkoyun201226} in Europe and Gamma-Ray Energy Tracking Array
(GRETA) in the US \cite{Lee20031095} are presently building such
arrays. Position sensitive High-Purity Germanium (HPGe) detectors will
cover close to 4$\pi$ of solid angle and track the path of the
$\gamma$ rays inside the detector medium giving maximum efficiency and
an excellent energy resolution. The technique of $\gamma$-ray tracking
allows both the high efficiency needed for high-fold coincidences and
the excellent position resolution needed for Doppler Correction at
in-flight fragmentation facilities.

Gamma-ray tracking starts from the digitally recorded wave-forms of
the pre-amplified signals of the highly-segmented HPGe detectors. The
wave-forms are treated with Pulse Shape Analysis (PSA) techniques to
extract the position of the interaction points of the $\gamma$ rays in
the detector, presently with a position resolution of about 5~mm FWHM
\cite{RECCHIA200960,RECCHIA2009555,Soderstrom201196}.  The interaction
points ({\it hits}) are grouped into events on the basis of their {\it
  timestamp}, {\it i.e.} the absolute time of the $\gamma$-ray
interaction. The sequence of interaction points of the $\gamma$ rays
in the same event is reconstructed from the hits via {\it tracking}
algorithms. A higher efficiency with a high peak-to-total is expected
as the solid angle taken by Anti-Compton shields is now occupied by
HPGe crystals and the Compton event suppression is performed by the
$\gamma$-ray tracking algorithm. The use of digital electronics allows
a higher count-rate with maintained energy resolution, and rates up to
50kHz per crystals are routinely used during experiments. The almost
continuous measurement of $\gamma$-ray emission angles, via the PSA
and tracking, allows for the excellent Doppler correction seen in
$\gamma$-ray tracking arrays and opens up a new degree of sensitivity
in the determination of nuclear structure observables such as
electromagnetic moments (e.g. lifetimes measurements based on Doppler
shift and perturbed angular correlations). This paper is meant as both
a snapshot in time of the capacities of AGATA and as a reference paper
to be used when analysing data from AGATA experiments performed at
GANIL.

The first experimental campaign with the demonstrator AGATA sub-array
was at LNL (2009-2011) \cite{Gadea201188} where it was coupled to the
PRISMA spectrometer for the study of neutron-rich nuclei produced in
fusion-fission and neutron-transfer reactions. This was followed by a
campaign at GSI (2012-2014). Here a larger AGATA sub-array was coupled
to the FRS separator \cite{Domingopardo2012297} for the first campaign
with radioactive ion beams. The performance of the AGATA sub-array at
GSI has been extensively studied \cite{Lalovic2016258}, with focus on
the efficiency of the AGATA sub-array as a function of energy and data
treatment. Other performance aspects such as the peak-to-total ratio
were also investigated.

Since 2015, AGATA has been operating \cite{Clement20171} at GANIL,
Caen, France, where it has been coupled to VAMOS (a variable mode high
acceptance spectrometer)
\cite{Savajols19991027c,Rejmund2011184}. Three campaigns of
measurements have been performed with focus mainly on neutron-rich
nuclei populated using multi-nucleon transfer reactions or via
fusion-fission or induced fission. In 2018, a campaign with AGATA
coupled to the NEDA \cite{VALIENTEDOBON201981} neutron detector and
the DIAMANT \cite{SCHEURER1997501,GAL2004502} charged particle
detector was performed. AGATA is foreseen to stay at GANIL until the
middle of 2021. A campaign of source measurements was performed during
2016 to, together with in-beam data, quantify the performance of AGATA
at the GANIL site as well. Basic performance data such as efficiencies
are needed to analyze the data taken during the campaigns, but a
careful follow-up of the evaluation of the AGATA performance as the
size of the array changes, detectors and electronics age and/or are
changed is also of considerable interest. It allows one to ensure that
the performance is in accordance with expectations. Furthermore, it
helps understand where efforts to improve are important - this both at
a fundamental level, e.g. Pulse-shape analyses or $\gamma$-ray
tracking, and on a more practical level learning how to best maintain
the system at a high level of performance. Extensive Monte Carlo
simulations of AGATA are performed as well in order to predict the
performance in different experimental configurations and with
different number of AGATA crystals. A thorough evaluation of the
performance of such a detection system allows for the bench-marking of
the Monte Carlo simulations, further helping the analysis of
experimental data.

There is an extensive literature on the performance of $\gamma$-ray
tracking arrays
(e.g. \cite{KORICHI201780,LAURITSEN2016,Wiessharr2017}) that address
the questions of efficiency, peak-to-total, and, Doppler correction
capabilities of $\gamma$-ray tracking arrays. As this paper aims at
giving a snap shot in time of AGATA and its capabilities during the
AGATA at GANIL campaign no detailed comparisons are made with the
literature as in most cases significant differences in setups and
methodology would require extensive discussion to make sense of such
comparisons.

In this paper we will describe the performance of AGATA as of mid
2016, when it was equipped with 30 crystals. In section \ref{sec:exp}
and section \ref{sec:dataprocessing} the experimental set up and data
acquisition are presented. The performance of individual crystals is
discussed in section \ref{sec:crysperf}. In section
\ref{sec:perfagata} the performance of AGATA as an array is discussed,
using the Orsay Forward Tracking algorithm.  Estimates of the position
resolution achievable in a typical experiment are given in section
\ref{sec:psares}.  As the angular coverage of AGATA increases the
capabilities in terms of measuring angular correlations increase and
this is discussed in section \ref{sec:angcorr}. Conclusions are given
in section \ref{sec:conclusions}.

\section{Experimental setup and data taking}
\label{sec:exp}
\begin{figure}[htb]
\centering
 \includegraphics[width=.5\textwidth]{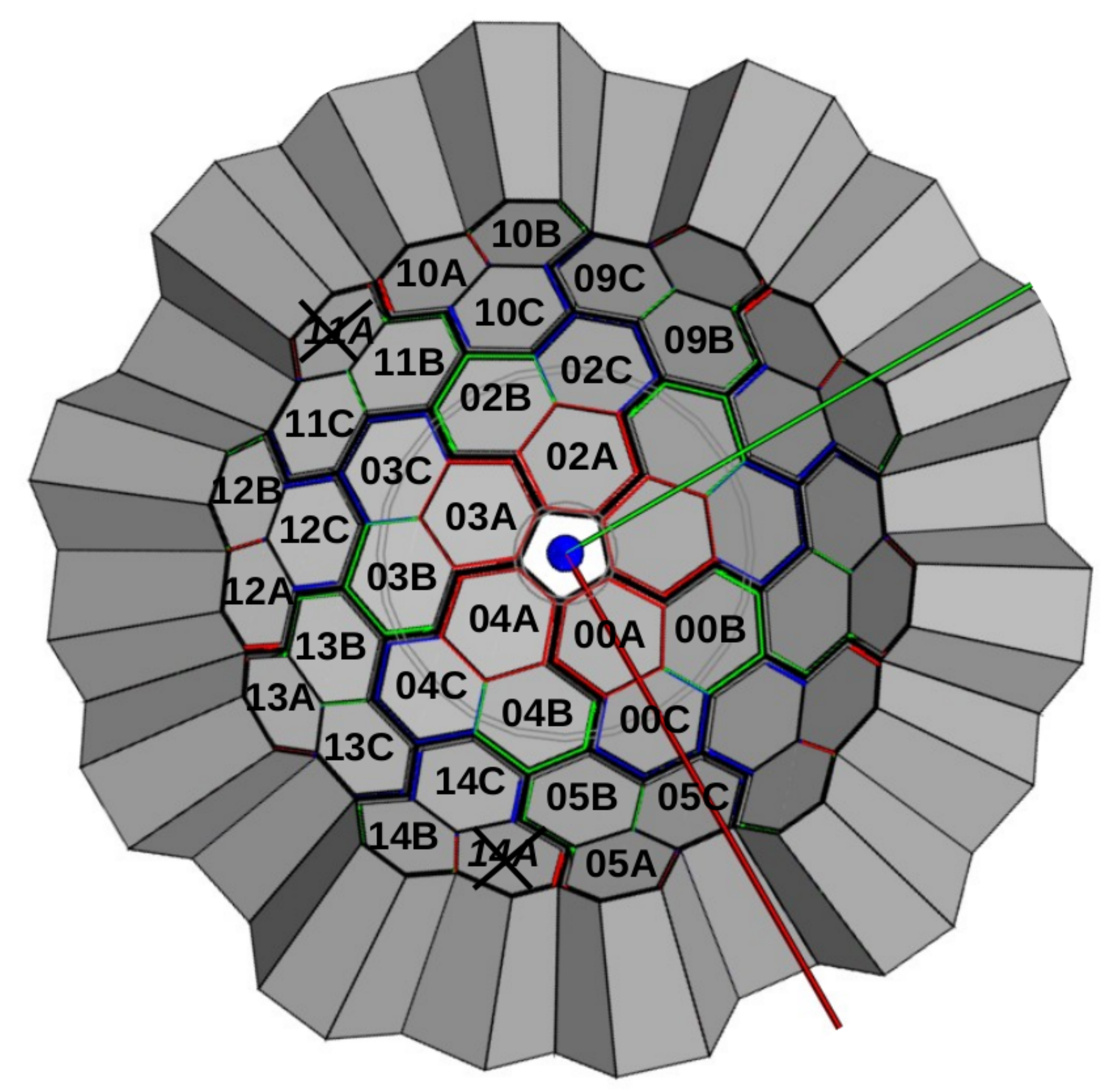}
 \caption{AGATA detectors seen from the reaction chamber point of
   view, labeled according to their position in the honeycomb. The two
   crossed over detectors are physically present, but not connected
   to an electronic channel. Positions with no labeling are
   empty. The red and green line are the x-, and, y-axis,
   respectively, for the installation in Legnaro and GSI, showing the
   rotation made of the structure at GANIL.\label{fig:labeling}}
\end{figure}

In 2016 the AGATA array consisted of 10 triple clusters (Agata Triple
Cluster, or ATC) \cite{Wiens2010223} and one double cluster (Agata
Double Cluster or ADC) arranged as schematically represented in figure
\ref{fig:labeling}. Two of the detectors present in the frame where
not connected to an electronics channel, giving a total of 30 active
detectors. One detector showed varying performance, related to the
electronics that was used, and is excluded form the efficiency
determinations. Measurements were performed both at what is referred
to as \lq\lq nominal position\rq\rq\, {\it i.e.} the front surfaces of
all AGATA crystals are positioned at 23.5 cm from the target position,
and at \lq\lq compact position\rq\rq\ with a distance of 13.5 cm
between the closest part of the imaginary sphere that touches the
front of the AGATA crystals and the target position.
Different standard radioactive sources were placed at the target
position, see Tab. \ref{tab:sources}. The Aluminum materials
surrounding the target position, {\it i.e.} reaction chamber and
target holder, were the same as in most of the experimental setups of
the campaign. These aluminum structures are included in the Geant4
simulations \cite{Agostinelli2003250,Farnea2010331} presented in this
work.
\begin{table}[tbh]
\centering
\begin{tabular}{|l|l|l|}
\hline
Source & Activity [kBq] & t$_0$ \\
\hline
$^{152}$Eu & 19.1  & 05/01/2016 \\
\hline
$^{60}$Co & 8.7  & 05/01/2016 \\
\hline
\end{tabular}
\caption{Radioactive sources used for the measurement}
\label{tab:sources}
\end{table}

For each detector, the data were collected from the 36 segments as
well as for two different gains of the central contact (ranges of
$\approx\,$8 and 20 MeV).  The segment signals are referred to with a
letter A-F and a number 1-6 where the letter gives the sector of the
crystal and the number the slice, {\it i.e.} the segmentation
orthogonal to the bore hole for the central contact. The AGATA raw
data for each crystal in an event consist of, for each segment and for
the central contact, the amplitude and 100 samples (10 ns time between
samples, $\approx$ 40 pre-trigger and $\approx$ 60 post-trigger) of
the rise-time of the waveform and a time-stamp, used for the event
building.  For the source data used in this paper the amplitude was
extracted from a trapezoidal filter with a shaping time of 10~$\mu$s
followed by a flat top of 1$\mu$s. The online and offline data
processing is done using the same computer codes, and are described in
detail in section \ref{sec:dataprocessing}.

The preamplifier outputs were digitized and pre-processed by two
different generations of electronics. The ATCA phase 0 electronics was
developed at an early stage of the project for the AGATA Demonstrator,
described in Ref. \cite{Akkoyun201226}. For the GANIL Phase a second
generation of electronics was developed referred to as the GGP's
\cite{Barrientos_2015}. The two generations of electronics use the
same algorithms for determining the energy. However, for the
determination of the time of a signal, the ATCA phase 0 electronics
use a digital constant fraction (CFD) whereas the GGPs use a low-level
leading edge algorithm. These times are used for triggering purposes
only. For both generations of the electronics discussed above digital
CFDs are used for proper timing when analysing the data. A time signal
is also extracted directly from the digitizer in order to provide a
$\gamma$-ray trigger for the VME electronics of VAMOS.

\section{Data processing}
\label{sec:dataprocessing}
\begin{figure*}[htb]
\centering
 \includegraphics[width=\textwidth]{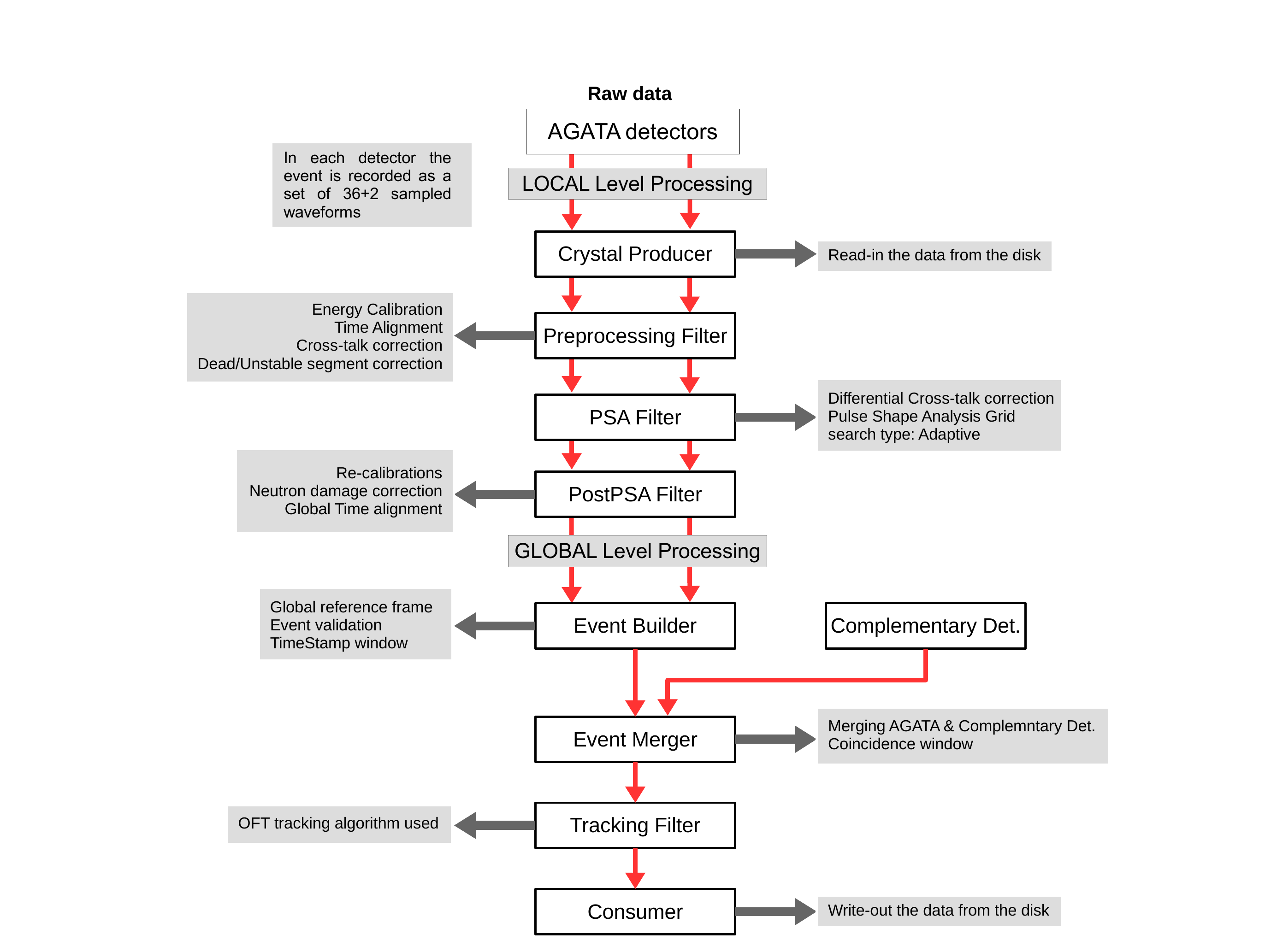}
 \caption{Chains of Narval actors used for data processing. For
   details see text. \label{fig:actors}}
\end{figure*}

The raw data (event-by-event amplitude, timestamp and traces for
segments and central contact) are treated with the chain of Narval
actors as depicted in figure \ref{fig:actors}. Starting at the top we
have data coming from the front-end electronics into the computer farm
with the first Narval actor \cite{Akkoyun201226}, the \lq\lq Crystal
Producer\rq\rq\ that puts the data of the crystal into the Agata Data
Flow. The next step, done in the \lq\lq Preprocessing Filter\rq\rq\ is
to perform energy calibrations, time alignments, cross-talk
corrections and the reconstruction of data in crystals, which are
missing a segment (in case of several missing segments this is no
longer possible), see section \ref{sec:crysperf}. Following the
preprocessing comes the pulse-shape analysis where the $\gamma$-ray
interaction positions are extracted using an adaptive grid search
algorithm~\cite{venturelli2004adaptive} where the experimental pulses
are compared to pulses calculated using the Agata Detector
Library~\cite{Bruyneel2016}. Tests allowing to search for more than
one interaction per segment of an AGATA crystal have been
performed. These tests have shown no improvement in terms of
efficiency and peak-to-total after $\gamma$-ray tracking so presently
the search is limited to one interaction per segment. From this point
on the traces are removed from the data flow. In a typical experiment
the result from the PSA are also written to disk at this point as this
allows redoing the subsequent steps in the analysis offline without
the time consuming PSA. The final step where the data from each
crystal is treated individually (Local Level Processing) is the \lq\lq
Post PSA\rq\rq, in which, apart from timestamp realignments, several
energy correction procedures described in section \ref{sec:crysperf}
are performed. After this, data from all AGATA crystals are merged in
the \lq\lq Event Builder\rq\rq\ on the basis of a coincidence
condition using the individual time stamps of each crystal. This is
the start of what is referred to as the Global Level
Processing. Complementary detectors are added into the Agata Data Flow
in the \lq\lq Event Merger\rq\rq. This is done before $\gamma$-ray
tracking because complementary data from these detector, e.g. data
from a beam tracking detector in case of a very large beam spot, are
potentially of use for the tracking of the $\gamma$ rays. Finally
$\gamma$-ray tracking is performed. In this work the OFT $\gamma$-ray
tracking algorithm has been used~\cite{Lopezmartens2004454}. Finally
the data is written to disk by a \lq\lq Consumer\rq\rq. This procedure
is performed online for monitoring of the experiments (data
processing) but also performed as a part of the data analysis (data
replay) starting from the raw traces or from the interaction points
given by the online PSA. The possibility to also store the
experimental traces to disk depends on the experimental conditions,
and is in practice only possible if the number of validated events is
lower than about 3 kHz per crystal (inducing a dead time of about
15\%).  Automatic procedures have been developed, both for energy
calibration purposes and for the preparation of the configuration
files that the actors use allowing error free and fast analyses of
experimental data.

\section{Crystal Performance}
\label{sec:crysperf}
In this section the performance in terms of energy and resolution for
each crystal is discussed, named with their position in the AGATA
frame at GANIL. The performance of the individual detectors was
determined using measurements with $^{60}$Co and $^{152}$Eu sources,
see table \ref{tab:sources}. A set of standard procedures are
performed to minimize the FWHM for each crystal. These procedures
consist of cross-talk corrections and neutron-damage correction. The
energies for events with more than one segment with net charge have to
be corrected for cross talk, mainly between the central contact and
the segments, as the energy calibration is performed mainly with
events with segment multiplicity 1. Correction coefficients are
extracted from source data either by looking at the shift of the
full-energy peak made by the summing of segments in fold two events or
by looking at the base-line shift in fold one events. This procedure
has been described in detail by Bruyneel at
al. \cite{Bruyneel2009196,BRUYNEEL200999}. The correction for the
effects of the neutron damage on the detection of the $\gamma$ rays of
interest has been performed following the theoretical approach
described by Bruyneel and coauthors \cite{Bruyneel2013}. Two
calibration coefficients for every detector channel, used to correct
for the electron and hole trapping, are determined. This is done using
a grid-search based minimization of the FWHM and the left tail of the
peaks in the spectra for each channel, {\it i.e.} 37 per detector.  In
figure \ref{fig:ndamage} the effect of the correction is shown for one
detector.  This correction is more important for the segments as they
are more sensitive to hole trapping, but it is also done for the
central contact, and it is thus important also when the sum energy of
hits inside a crystal for an event is normalised to the value measured
by the central contact. This correction is particularly important for
measurements of lifetimes via line-shape analysis techniques, where
the symmetry of the detector response function is extremely important
to minimize systematical errors in the lifetime determination.

\subsection{Energy resolution}
\label{subsec:crysenergyres}
\begin{figure}[htb]
\centering
 \includegraphics[width=.5\textwidth]{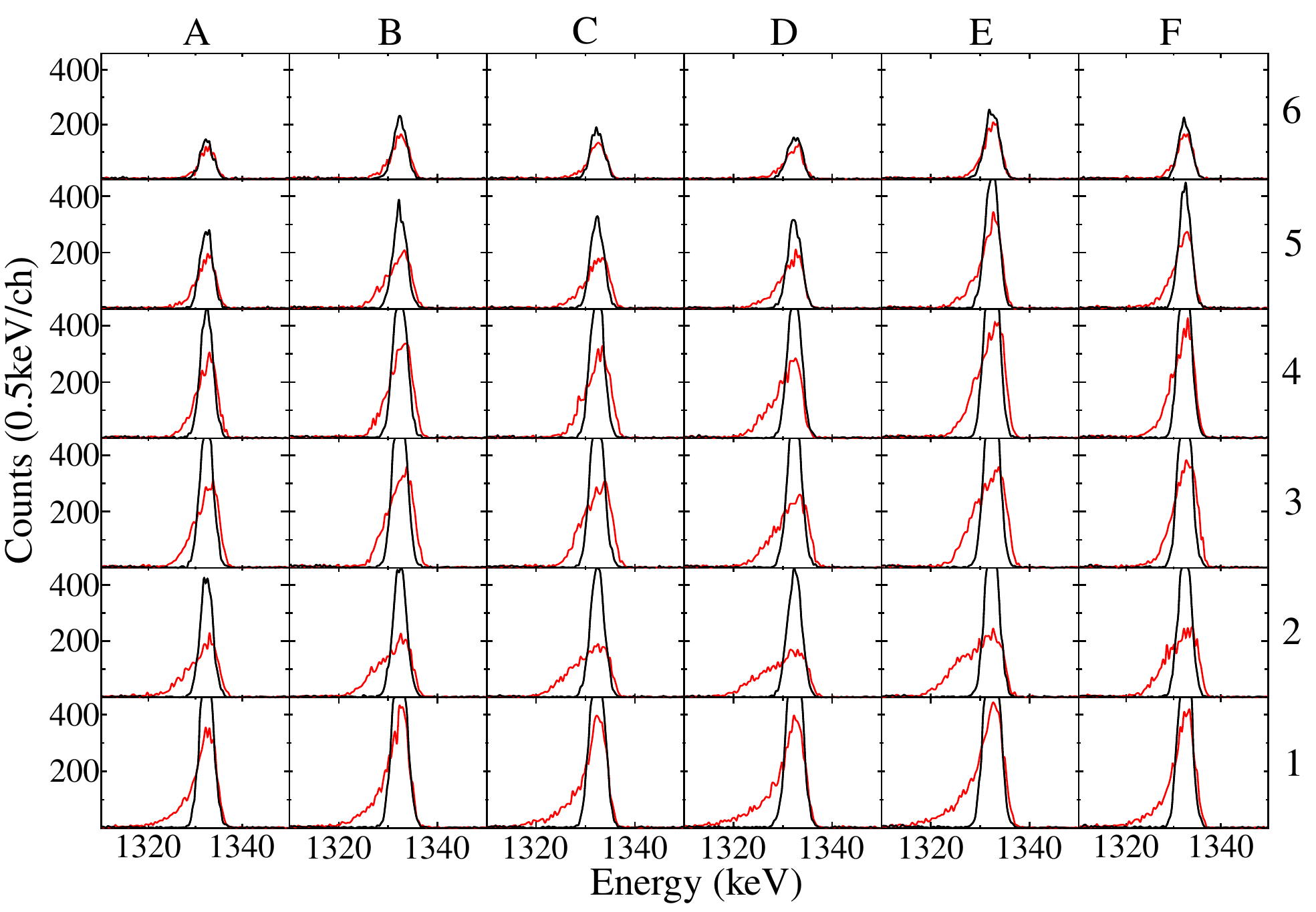}
 \caption{ \label{fig:ndamage} Example of the peak line shapes for the
   1332 keV 60Co $\gamma$-ray before (red) and after (black) the
   neutron damage correction for the 36 segments of the crystal A002
   position 12A ATC3. }
\end{figure}

The energy resolution has been determined for each segment and central
contact for the crystals in the array at the moment of taking source
data (2016). After the exposure to fast neutrons produced in deep
inelastic collisions, fission and fusion evaporation reactions in the
first campaign at GANIL in 2015, several AGATA crystals were damaged
by the charge traps created by neutron radiation damage in the Ge
crystal. For the most exposed detectors the integrated flux exceeds
$10^9$ n/cm$^2$. This is based on the deterioration of the uncorrected
FWHM \cite{Ross2009}. These traps are lattice defects that lead to a
reduction of the charge collection efficiency which appears as a low
energy tailing on the energy line shape (red line in figure
\ref{fig:ndamage}). In position sensitive Ge detectors, like the AGATA
ones, it is possible to apply an empirical correction to the neutron
damage effects \cite{Bruyneel2013}. These corrections were applied to
20 of the AGATA crystals in this work (see table \ref{table:DetId} in
appendix \ref{app:detid}). As an example of the effect of neutron
damage correction the original spectra and the ones after the
corrections, for one detector, are shown in figure
\ref{fig:ndamage}. For the other 10 detectors good energy resolution
was achieved without the neutron-damage correction procedure.

\begin{figure}[h]
  \centering
  \includegraphics[width=.5\textwidth]{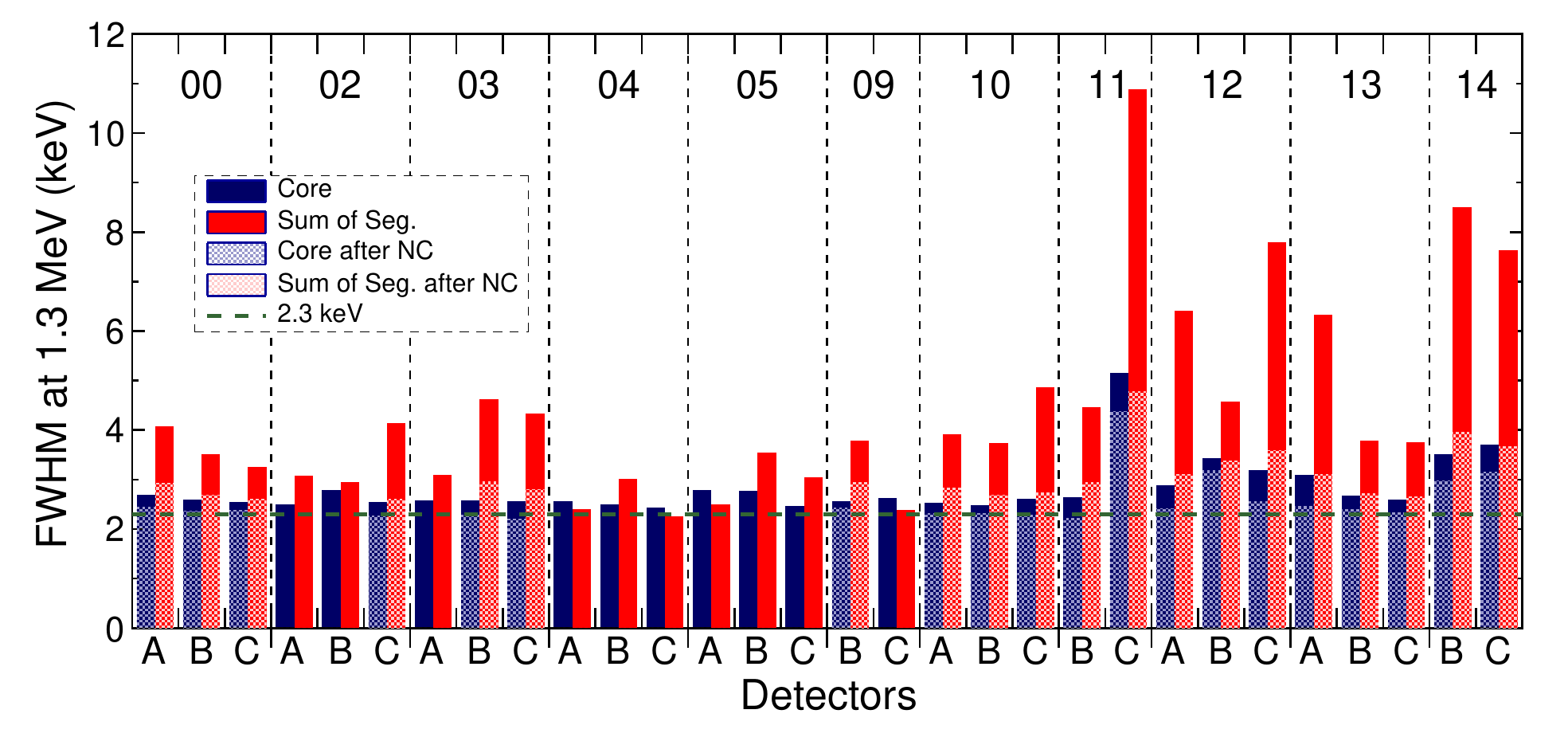}
  \caption{FWHM at 1.332 MeV ($^{60}$Co) of the central contact (blue)
    and the sum of segments (red) before (dark colours) and after
    (light colours) the neutron damage correction for 20 out of the 30
    capsules individually named with its position labels. }
	\label{fig:resol}
\end{figure}

\begin{figure}[h]
  \centering
  \includegraphics[width=.5\textwidth]{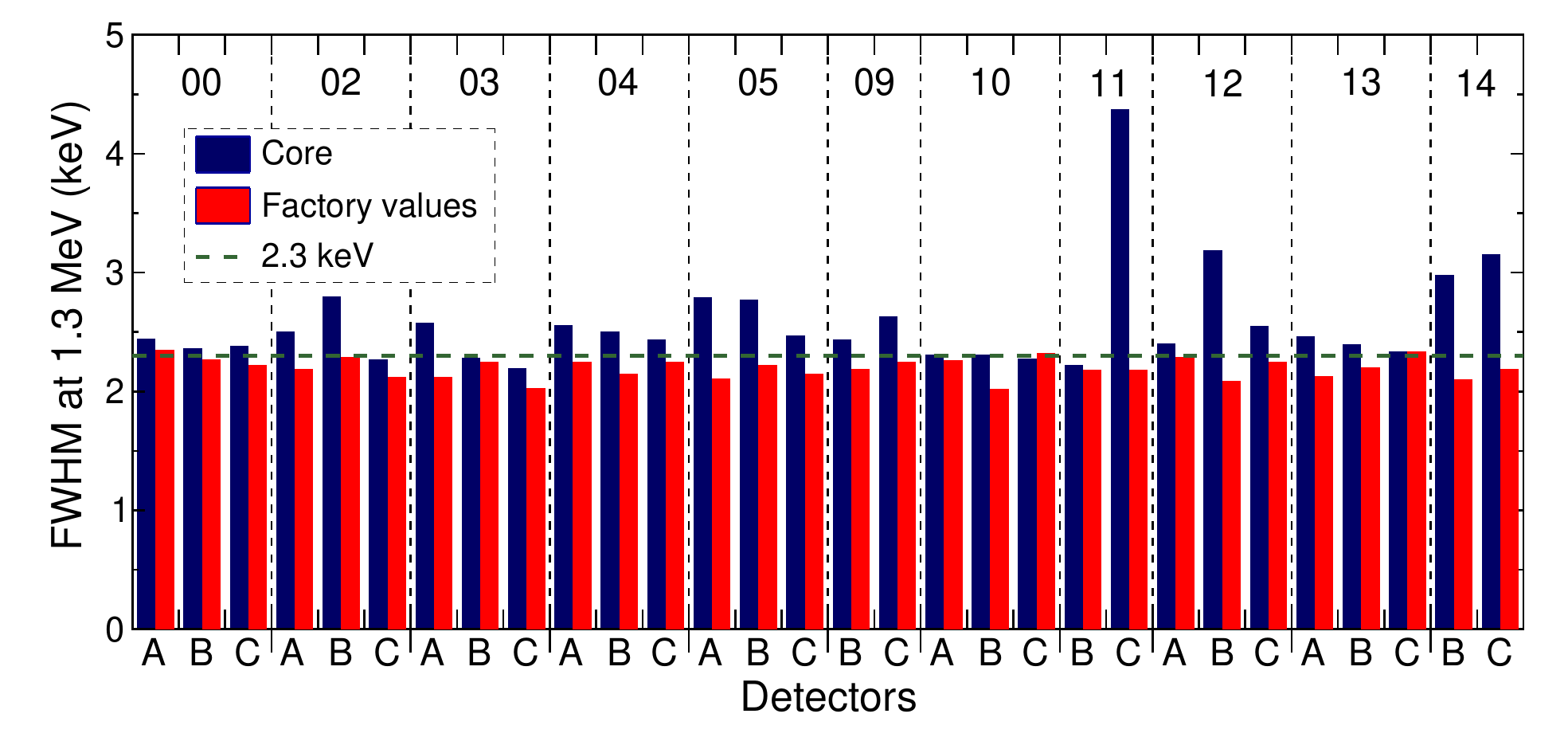}
  \caption{FWHM at 1.332 MeV ($^{60}$Co) of the central contact (blue)
    after the neutron damage correction compared with the original
    FWHM measured by Canberra (red) for the 30 capsules individually
    named with its position labels.}
	\label{fig:resolcan}
\end{figure}

In figure \ref{fig:resol} the resolutions for the central contacts and
sum of segments for the used detectors are reported. The average FWHM
resolution found for the central contacts before the neutron damage
correction is 2.93 keV and is improved to 2.57 keV after
correction. In the case of the sum of segments the average FWHM is
improved from 5.22 keV to 3.08 keV, showing the difference in
sensitivity to charge trapping. The comparison with the resolutions
taken from detector data sheets or factory measurements is reported in
figure \ref{fig:resolcan}. In general all the measured FWHM
resolutions for the crystals agree with the original ones, except for
the detector 11C (B013), which apart from being neutron damaged had a
resolution problem during the measurements, in both central contact
and segments, due to problems with the electronics.

\subsection{Crystal efficiency}
\label{subsec:cryseff}
The efficiency of each crystal has been determined first from the
central contact signal. Although this is not the normal operation mode
when performing $\gamma$-ray spectroscopy with AGATA, the crystal
central contact efficiency allows easier diagnostic of the Data
Acquisition Chain and easier comparison to Geant4 simulations. For
these reasons, it is of great value. Two sets of data for efficiency
measurement at the nominal and compact position of AGATA have been
taken. All efficiency numbers quoted in this section are corrected for
dead time of the data acquisition system. 

The efficiency has been determined both from $\gamma-\gamma$
coincidences, corrected for the angular correlation effects for the
given geometry, and from singles central contact data taken with
$^{60}$Co and $^{152}$Eu sources. The coincidence data are not
affected by dead-time of the processing chain. The singles central
contact measurement is. To bypass this effect, the latter have been
recorded in coincidence with the VME/VXI electronic of the GANIL
acquisition system coupled to AGATA via the AGAVA board
\cite{Akkoyun201226,Clement20171}. The GANIL acquisition system is
triggered by the OR of the AGATA digitizers CFDs, triggering the AGAVA
board. The individual AGATA channels are then validated by the AGAVA
request within a 300 ns coincidence time window. At the source rate,
the VME/VXI GANIL electronic has a dead-time of $40\mu s$ per read-out
event, greater than the AGATA electronic system, and it can be
precisely quantified and used for live time correction in the single
central contact efficiency measurement. For the $\gamma-\gamma$
coincidence, the 1332 keV-1173 keV from the $^{60}$Co source and 121.8
keV-1408 keV, 121.8 keV-244.7 keV and 344.3 keV-778.9 keV coincidences
from the $^{152}$Eu source were used. For fitting the $\gamma$-ray
peak areas used to extract the efficiency values, the Radware software
package was used \cite{Radware}. A background subtraction was made by
evaluating the correlated background on both sides of the gating
energy for the $\gamma-\gamma$ coincidences analysis.

\begin{figure}[h]
  \centering
  \includegraphics[width=.5\textwidth]{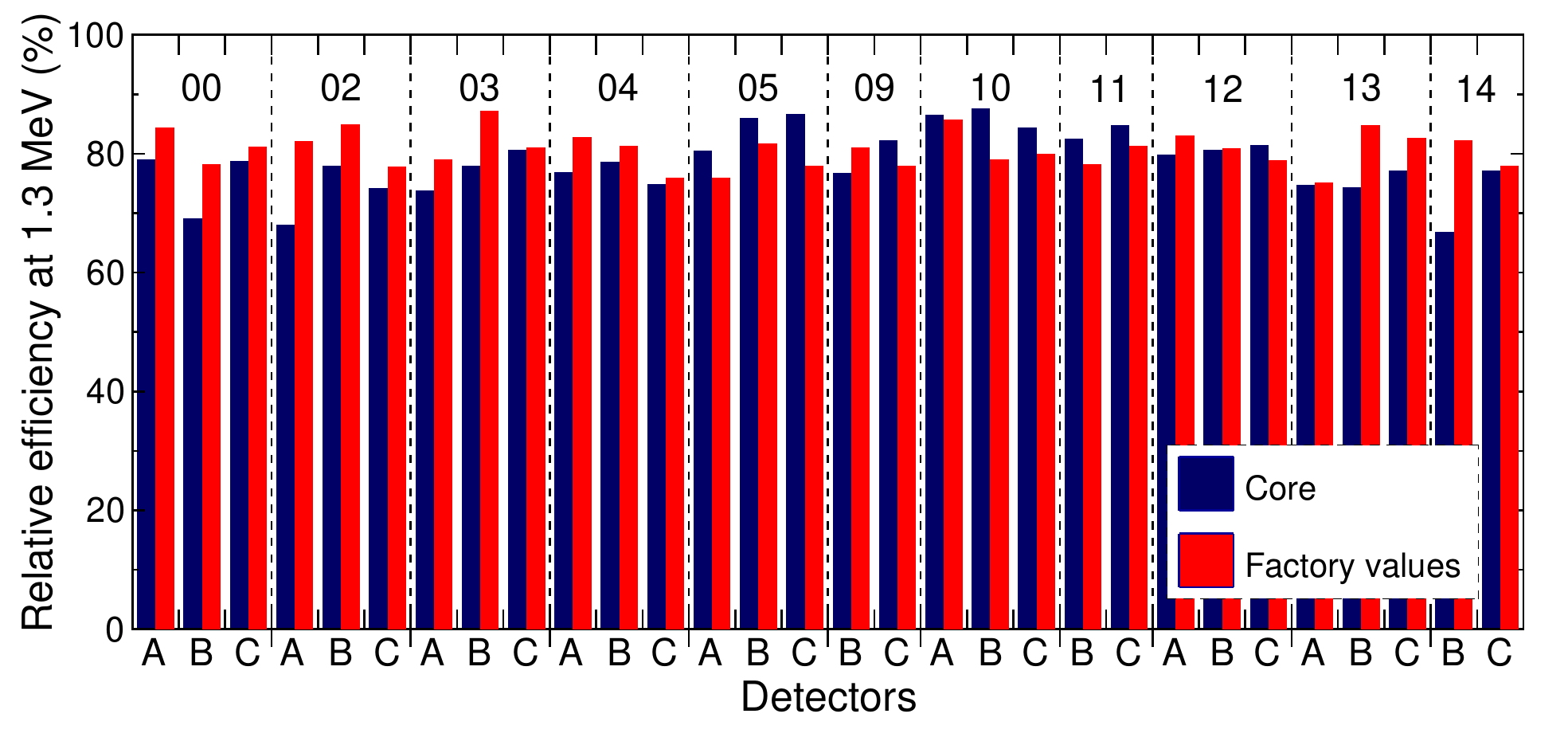}
  \caption{Relative central contact efficiency at 1.3 MeV ($^{60}$Co)
    in comparison with the initial relative efficiency as provided by
    manufacturer for the 29 capsules individually named with its
    position label.}
  \label{fig:eff_cry}
\end{figure}

Using the 1.3 MeV transition from the decay of $^{60}$Co the
efficiency relative to a 3 in $\times$ 3 in NaI detector ({\it i.e.},
$1.2×10^{-3}$ cps/Bq at 25 cm) for each detector at nominal position
was extracted and is reported in figure \ref{fig:eff_cry}. In the same
picture, the value at 1.3 MeV as measured at the factory or during the
customer acceptance tests is shown.  The average measured value is
$79$\% with a sample standard deviation of $5$\%, close to the factory
values average of $81$\%. Crystal 02C suffered from oscillations
during the measurement and is therefore excluded in the efficiency
numbers discussed below.

\begin{figure}[h]
  \centering
  \includegraphics[width=.5\textwidth]{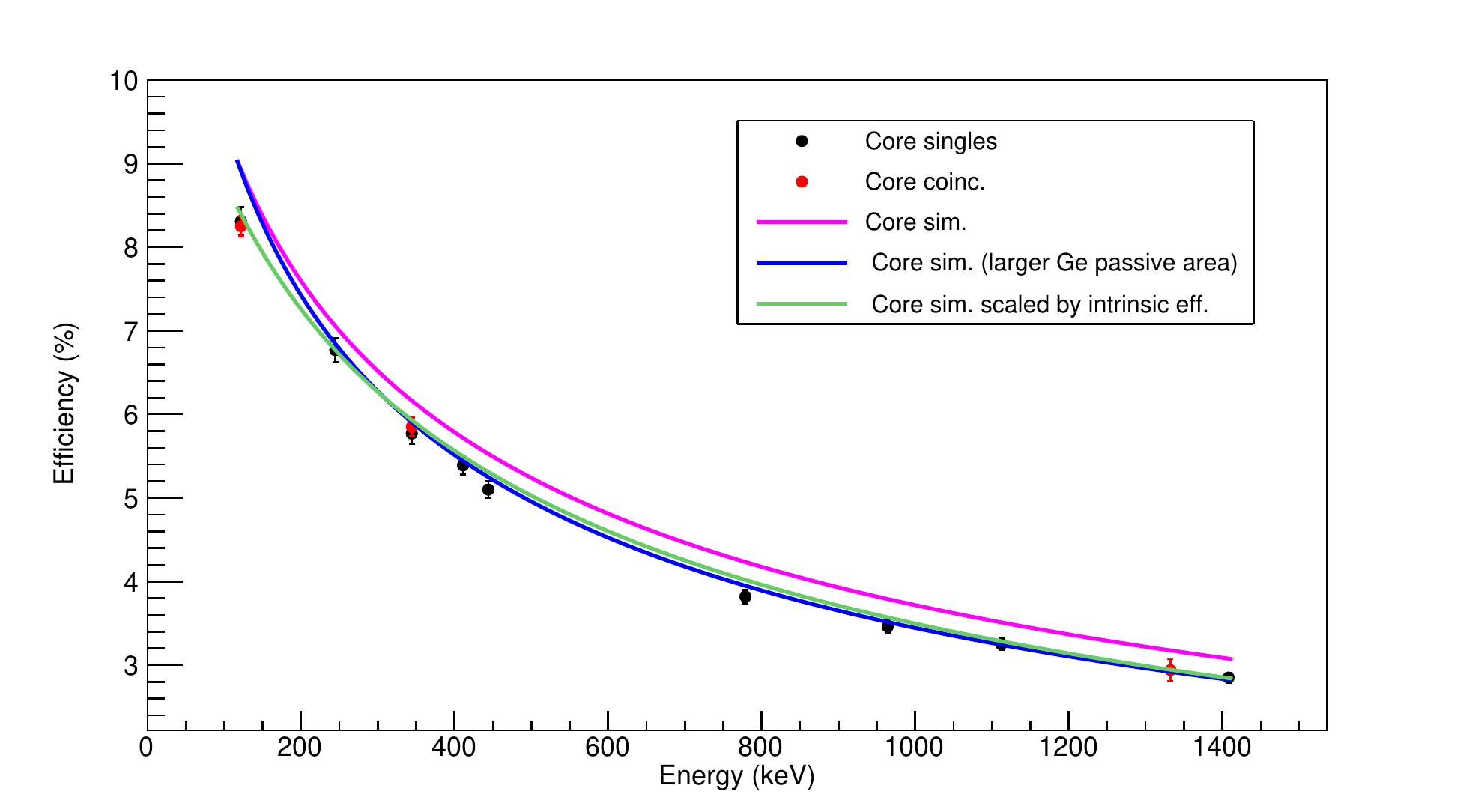}
  \caption{Absolute central contact efficiency for the 29 capsules
    AGATA sub-array in \textbf{nominal position} (23.5 cm distance
    source to detectors) obtained with the data from $^{152}$Eu and
    $^{60}$Co in singles (black circles) and in coincidences (red
    circles) in comparison with the simulations (magenta, blue, and
    green lines). The green line corresponds to simulations where the
    efficiency has been scaled according to the difference between the
    simulated absolute efficiency and the measured absolute
    efficiency. Simulations performed with increased dead layers are
    also shown (blue line). See text for details. The rate
    per crystal at this position was around 200 Hz.}
  \label{fig:nominal}
\end{figure}

\begin{figure}[h]
  \centering
  \includegraphics[width=.5\textwidth]{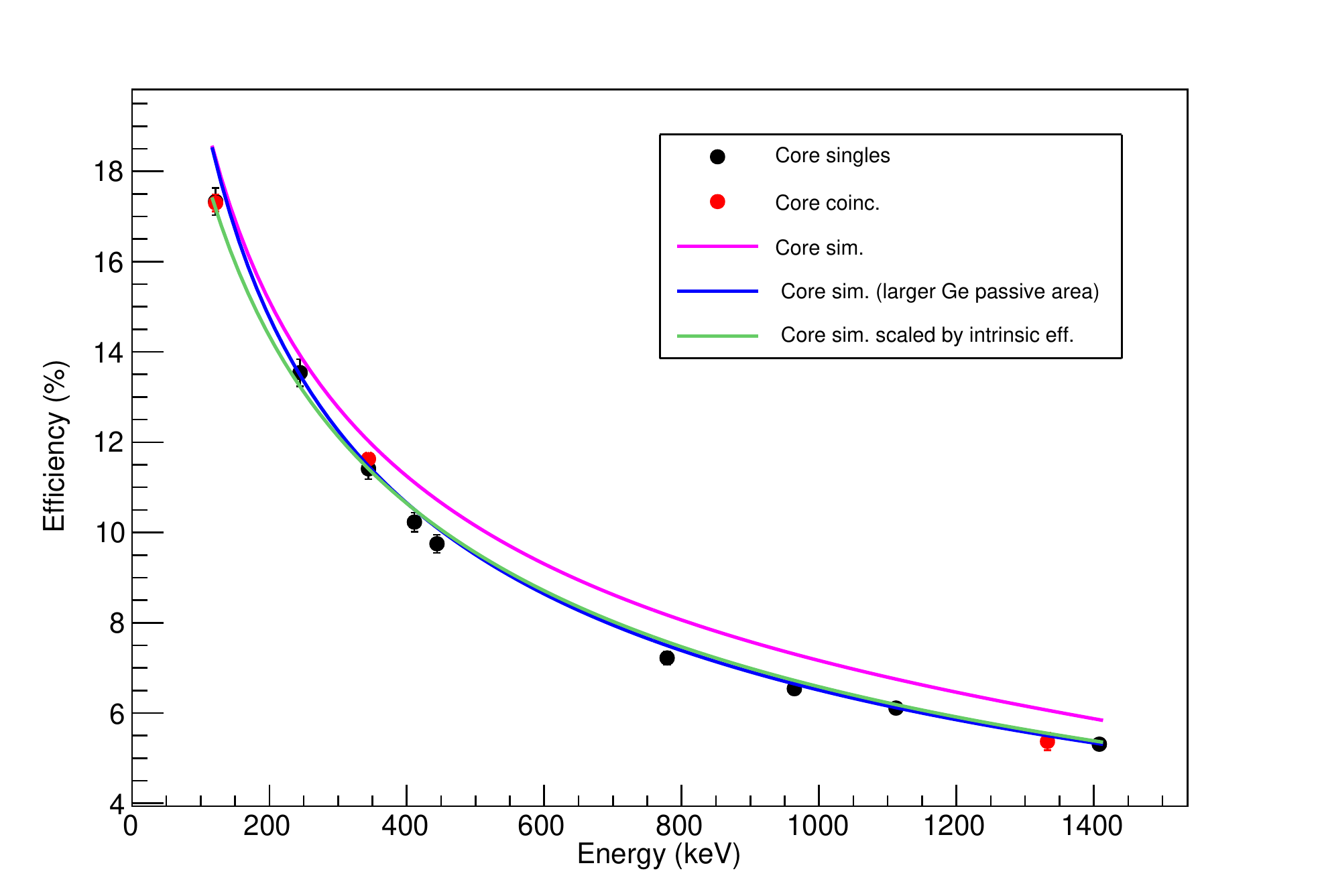}
  \caption{Same as figure \ref{fig:nominal} for the compact
    configuration and with a data rate per between 300 and 500 Hz.}
	\label{fig:compact}
\end{figure}

The absolute central contact efficiency for the whole array, composed
of 29 operational crystals, is reported in figure \ref{fig:nominal}
for the nominal position of AGATA and figure \ref{fig:compact} for the
compact position of AGATA. Here each crystal is treated as a single
detector like in a standard $\gamma$-ray detector array. The values
obtained in the singles measurement are compared with the
$\gamma-\gamma$ results and simulations and overlap well. For the
nominal geometry, the efficiency measured using singles is 2.95(6)\%
at 1332 keV whereas for the compact geometry it is 5.5(1)\% at 1332
keV. Geant4 simulations using the AGATA simulation package
\cite{Farnea2010331} have been performed. These simulations include a
realistic implementation of the reaction chamber used during the
experimental campaign at GANIL \cite{Clement20171}, a steel block to
emulate the effect of the VAMOS quadrupole, as well as the two
crystals that were present but not used during the measurements. There
is 12\% discrepancy between the simulation and the experimental
results. This difference is larger than the 2.5\% average discrepancy
for the individual crystals, as shown in figure \ref{fig:eff_cry},
between factory measurements and measurements made within the AGATA
collaboration. However, Geant4 simulations of the three differently
shapes crystals used by AGATA give a relative efficiency of 86\%,
86\%, and 87\%, for type A,B, and C, respectively. The average
measured value is 79\%, or 8\% lower. This is in reasonable agreement
with the 12\% of efficiency missing when compared with simulations, as
is illustrated in figures \ref{fig:nominal} and \ref{fig:compact}
(green line) where the efficiency of each crystal has been scaled to
its measured value. Here the question of how the 12\% of effective
germanium is lost has to be raised. The presence of a dead layer or
missing germanium will have an impact on the PSA as the pulse shapes
depend on the active volume and shape of the germanium
diode. Simulations assuming a thicker dear layers improve the
correspondence with experimental data. It is however difficult to pin
down the contribution from different surfaces of the detectors, {\it
  i.e.}, one can reproduce experimental data with different
combination of dead layers around the central contact and at the back
of the detector. Moreover, the mismatch of the efficiencies at low
energies cannot be corrected reducing the active volume around the
central contact or at the back of the detector. In figures
\ref{fig:nominal} and \ref{fig:compact} simulations with a dead layer
of 2.5 mm around the central contact and 3 mm at the back are also
shown. For estimates of dead layers in HPGe detectors see, e.g., the
work of Eberth and Simpson \cite{Eberth2008283} or Utsunomiya et
al. \cite{Utsunomiya2005455}

\section{Performance of the AGATA array with the Orsay Forward Tracking}
\label{sec:perfagata}

\paragraph{Description of OFT}
The Orsay Forward Tracking (OFT) algorithm \cite{Lopezmartens2004454}
was developed with simulated data sets produced with the Geant4 AGATA
code \cite{Farnea2010331}. The output of the simulations was modified
to emulate the expected experimental conditions, such as energy
resolution and threshold and position resolution allowing the
optimization of the algorithm using a realistic input. As all
forward-tracking algorithms the OFT starts with clustering interaction
points. These clusters are evaluated using a $\chi^2$-like test where
scattered energies after every interaction point as given by the
energies in each interaction point are compared to scattered energies
as given by the Compton scattering formula using the measured
positions of the interaction points. The best permutation for each
cluster is calculated and the clusters are sorted in order of best
figure of merit. Clusters that pass a threshold called $P_{track}$ are
accepted as good $\gamma$ rays. The most influential parameter in this
is $\sigma_{\theta}$ corresponding to the error in scattered energy
derived from the error in interaction positions from the PSA. Using
simulations this parameter was optimized to $\sigma_{\theta}$=2.4 mm
corresponding to the assumed position resolution in the simulations of
5 mm FWHM at 100 keV interaction point energy. Single interaction
points that are further away than 40 mm from the closest other
interaction point are treated as a photo-electric absorption
event. Here the probality for a $\gamma$ ray to have penetrated to a
given depth and been absorbed via the photo electric effect is
evaluated and compared to the $P_{sing}$ parameter.  The
single-interaction-point evaluation is an important part of the
tracking algorithm since the efficiency loss when it is not included
is very large for low-energy events, and non negligible at higher
energies: $\sim$20$\%$ of 1.4 MeV total-absorption events in each
individual detector are single interaction points. This last fact is
due to the way PSA identifies interaction points in the AGATA
detectors. As mentioned above, the Grid Search algorithm
\cite{venturelli2004adaptive} used online only looks for 1 interaction
point per segment. This is at variance with what is currently done at
GRETINA \cite{Paschalis201344} where the fits of the segment traces
allow for more than one hit per segment. For a detailed explatation on
the OFT algorithms see Lopez-Martens et
al. \cite{Lopezmartens2004454}.

\paragraph{OFT parameters}
The definition and typical ranges of the main parameters of OFT are
summarised in table \ref{parameters}.

\begin{table}[htp]
\begin{center}
  \caption{Table summarising the meaning and standard ranges of the
    main adjustable parameters of OFT. }
\label{parameters}
\begin{tabular}{|c|c|c|}
\hline
parameter & definition & typical value \\
\hline
$\sigma_{\theta}$ & average interaction-point  & 0.3-3 \\
 & position resolution (cm) & \\
\hline
$P_{sing}$ & minimum probability to accept  & 0.02-0.15 \\
 & single-interaction-point clusters &  \\
\hline
$P_{track}$ & minimum figure of merit to accept & 0.02-0.05 \\
 & mutliple-interaction-point clusters  &  \\
\hline
\end{tabular}
\end{center}
\end{table}%

Tuning the parameters can affect the spectral quality and shape.  As
an example, a high value of $\sigma_{\theta}$ corresponds to nearly
fully relaxing the comparison between scattered energies obtained from
interaction positions and scattered energies obtained from energy
differences. Basically, using a very large $\sigma_{\theta}$ reduces
the cluster evaluation stage to finding the most likely sequence of
interaction points in a cluster on the basis of ranges and interaction
probabilities only. Increasing $\sigma_{\theta}$ increases the
high-energy efficiency. However, it also decreases the low-energy
efficiency in the case of medium to high photon-multiplicity events
since single-interaction points are being accepted as members of
multi-interaction point clusters and are therefore lost as potential
$\gamma$ rays absorbed in a single interaction. There is an optimal
value of $\sigma_{\theta}$, which maximises the gain in efficiency at
medium and high energy while minimizing the loss of efficiency at low
energy. By analysing source and in-beam data obtained at Legnaro, GSI
and GANIL, the optimal value of $\sigma_{\theta}$ is found to be
around $\sim$6 and 8 mm. This corresponds to an average experimental
position resolution a factor of 2 to 3 worse than anticipated. This is
consistent with measurements of the position resolution of an
interaction point as a function of the deposited energy
\cite{Soderstrom201196} as well as with the observed clusterisation of
interaction points in specific areas of the detector segments.

Another example is given by the energy range of the single-interaction
spectrum, which grows when the threshold for validation of the
corresponding clusters is lowered. For $P_{sing}$=0.15, the spectrum
extends to $\sim$600 keV, while for $P_{sing}$=0.02, it goes beyond 2
MeV. Extending the spectrum increases the overall efficiency at
high-energy. There is however a trade off in the form of a larger
background: for $P_{sing}$=0.02, the single-interaction points are
responsible for nearly two thirds of the background present in the
spectrum of tracked photon energies.  Recent developments in the OFT
code have improved on this point by using an empirically deduced
energy-\lq\lq distance in germanium\rq\rq\ relationship instead of the
single parameter $P_{sing}$, allowing an improved peak-to-total. The
new single-interaction treatment, that is tailored not to have a
negative impact on efficiency, is further discussed in section
\ref{subsec:trackingpeaktototal}.

The optimal value of $P_{track}$ is found to be around 0.05. Some very
slight adjustments can be made as a function of $\sigma_{\theta}$, but
the general trend is that a smaller value leads to more background and
a larger value reduces the peak intensities.

\subsection{Tracking Efficiency measurements}
\label{subsec:trackingeff}
\begin{figure}[htbp]
   \centering
   \includegraphics [width=.5\textwidth] {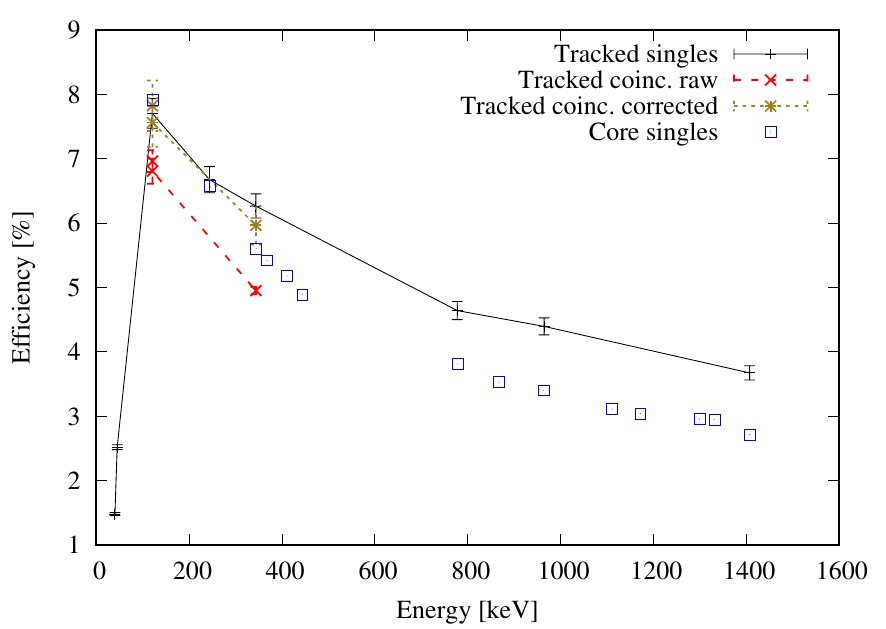} 
   \caption{Tracking efficiency of 29 AGATA detectors as a function of
     photon energy obtained with the standard OFT parameter set and
     using either the total singles tracked spectrum or the (121
     keV-244 keV), (121 keV - 1408 keV) and (344 keV - 778 keV)
     $\gamma$-$\gamma$ coincidences. The efficiency for 29 cores
     scaled from figure \ref{fig:nominal} is also shown. See text for
     details.}
   \label{Trackeff}
\end{figure}

The standard set of OFT parameters ($\sigma_{\theta}$=0.8,
{\it{P$_{track}$}} =0.05 and {\it{P$_{sing}$}} =0.05) were used to
extract the tracking efficiency of AGATA at GANIL in a configuration
with 29 capsules. The efficiencies to track the photons emitted by a
$^{152}$Eu source were obtained by comparing the detected peak areas
to the expected intensities given the source activity, the measurement
time interval and the electronics dead time. Since there are several
2-photon cascades in the radioactive decay of $^{152}$Eu, the
efficiencies at certain photon energies can also be measured by
comparing the detected peak area of a transition when a coincidence
with the transition of interest is required or not. The advantage of
this second method is that no knowledge of the source activity or dead
time of the system is required. The efficiencies obtained are shown in
figure \ref{Trackeff}.

The efficiency to track a 1.4 MeV photon with 29 capsules is found to
be 3.67(1)$\%$. This corresponds to an add back factor with respect to
the efficiency of the 29 detectors taken individually of 1.285(4).

\begin{figure}[htbp]
   \centering
   \includegraphics[width=.5\textwidth]
   {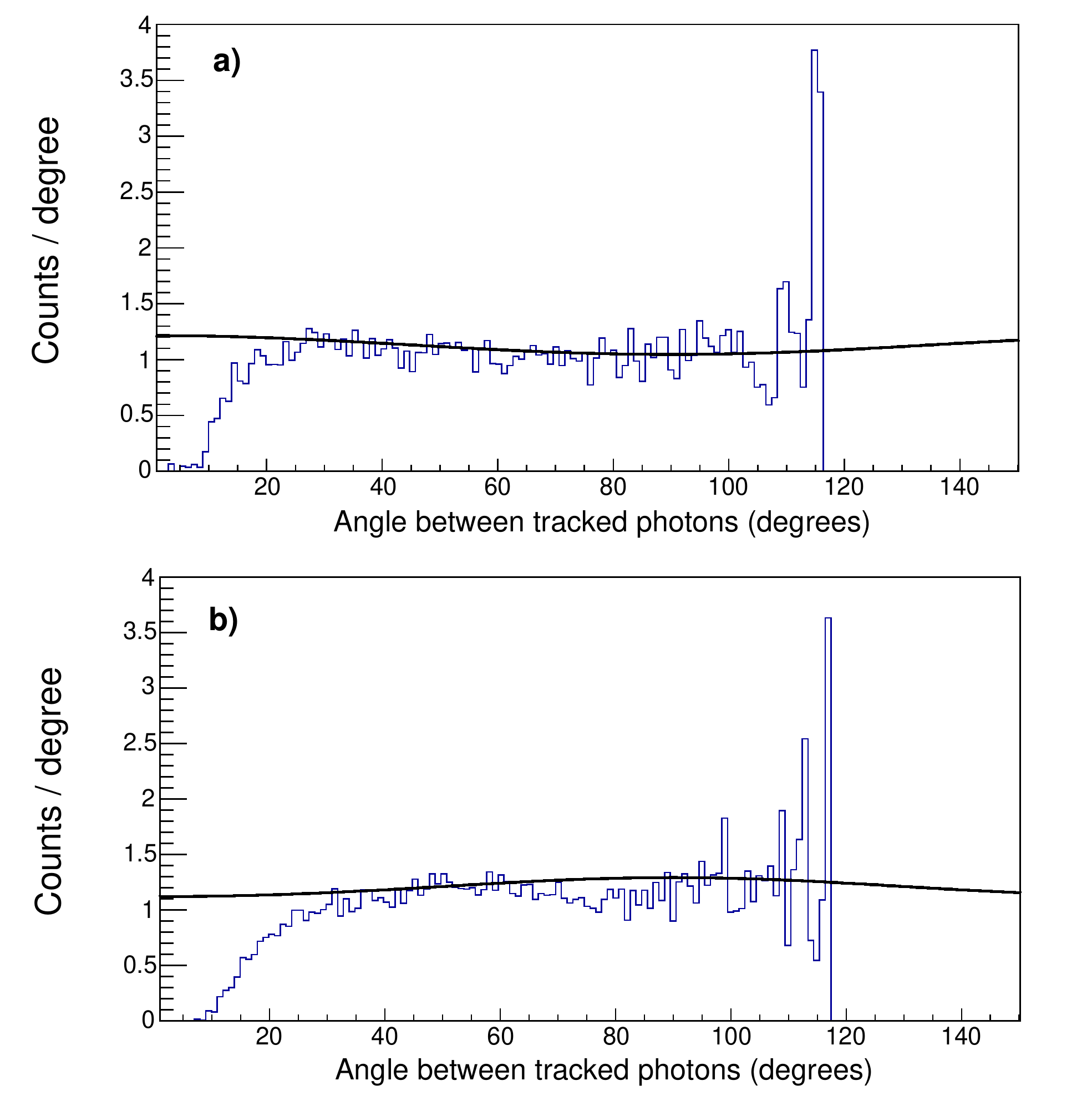} 
   \caption{a) $\gamma$-$\gamma$ angular correlations obtained for the
     121.8 keV - 244.7 keV cascade in $^{152}$Sm using the OFT
     parameter $\sigma_{\theta}$=0.8. b) same as a) for the 344.3 keV
     -778.9 keV cascade in $^{152}$Gd and in the case of
     $\sigma_{\theta}$=2.0. The solid lines represent the best
     adjustment of the theoretical curves to the data. The peaks at
     high angles are associated with large statistical errors not
     shown in order to keep the figure clear.}

   \label{AngularCorrelations}
\end{figure}

In figure \ref{Trackeff}, the raw coincidence efficiencies at 121 and
344 keV lie below the singles tracking efficiency curve. This is
because the tracking efficiency varies with the angle between the
emitted photons; most notably it vanishes for small angles due to the
deficiencies of the AGATA PSA algorithm and/or due to the fact that
the tracking algorithm cannot disentangle the points belonging to the
2 coincident photons when these lie too close to each other. This is
clearly seen in the plot of the $\gamma$-$\gamma$ angular correlations
for the 121.8-244.7 and 344.3-778.9 coincidences in $^{152}$Sm and
$^{152}$Gd shown in figure \ref{AngularCorrelations}. By correcting
the coincidence efficiencies by the missing fraction of the
experimental angular correlations compared to the theoretical curve,
the correct tracking efficiency values are recovered.

Using a larger value of $\sigma_{\theta}$ leads to a slightly lower
tracking efficiency below 200 keV, but yields 13$\%$ more efficiency at
1.4 MeV, making the add back factor increase to $\sim$1.4. It also
changes the raw coincidence efficiencies for some coincidence
couples. In the case of the 344.3 keV -778.9 keV cascade of
figure \ref{AngularCorrelations}, in particular, correlations are not
only absent at small angles, but also at larger angles, when OFT most
probably misinterprets all or a subset of the interaction points of
the event as points belonging to a back-scatter sequence.

\subsection{Tracking Peak-to-Total ratio}
\label{subsec:trackingpeaktototal}
\begin{figure}[htbp]
  \centering
  \includegraphics [width=.5\textwidth] {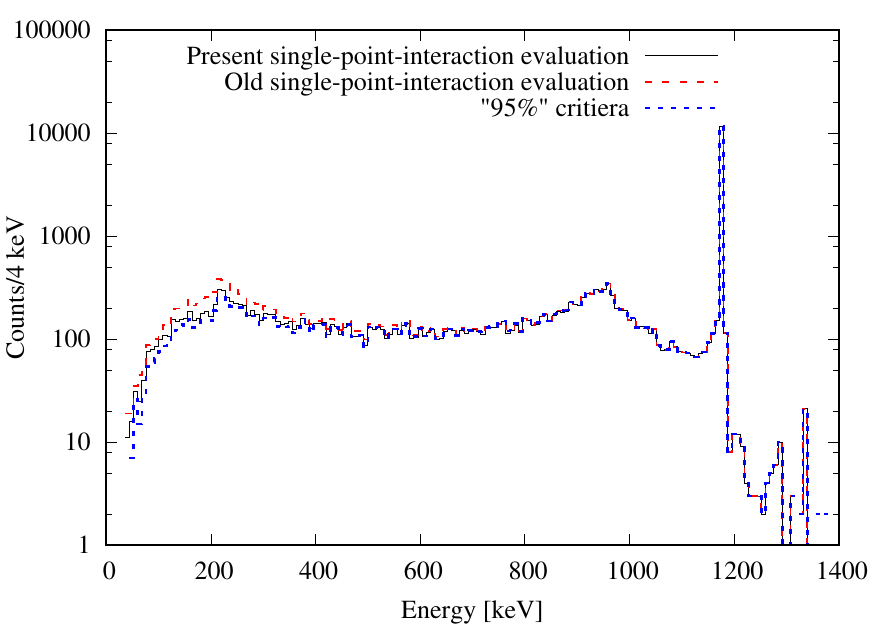}
  \caption{Gamma-ray tracked spectra for 29 AGATA detectors for
    $^{60}$Co data using a 30th AGATA detector as a external trigger
    by demanding the full absorption of the 1332.5 keV gamma in
    it. The solid line (black) is using the latest single-point
    interaction validation procedure, the dashed line (red) is using
    the old single-point interaction validation procedure, and finally
    the dotted line (blue) using the 95\% absorption limit.}
  \label{fig:co60gated}
\end{figure}

\begin{figure}[htbp]
  \centering
  \includegraphics [width=.5\textwidth] {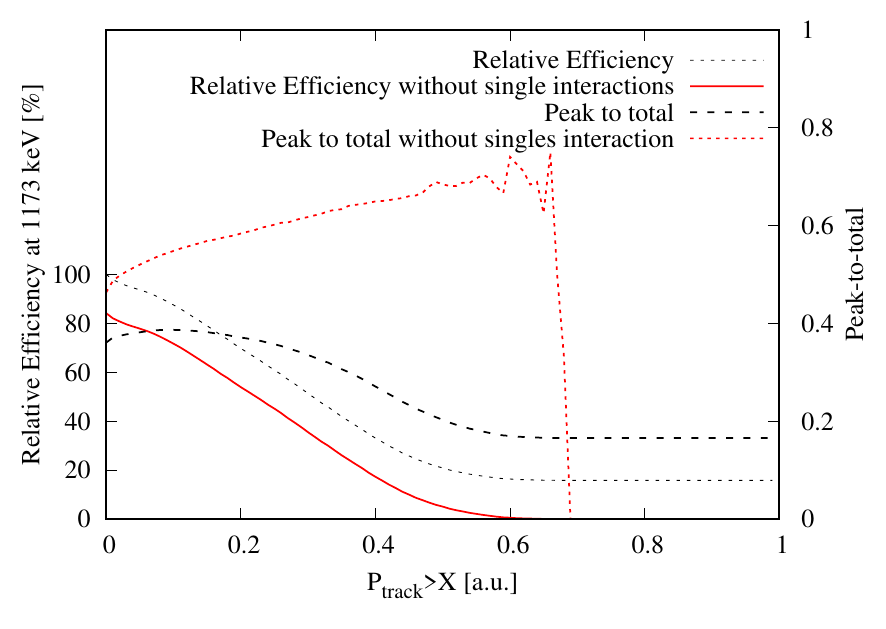}
  \caption{Efficiency relative to $P_{track}=0$ at 1173 keV and peak
    to total as a function of $P_{track}$ used by the OFT algorithm to
    accept or not a $\gamma$-ray track. This for when including or
    excluding single point interactions. }
  \label{fig:peaktototal}
\end{figure}

An important performance parameter for a $\gamma$-ray spectrometer is
the peak to total ratio quantifying the fraction of events found in
the full energy peak as compared to the total number of detected
$\gamma$ rays. Data was taken with a $^{60}$Co source with an activity
of $8.7$ kBq. Gamma-ray tracking was then performed offline for 29 of
the 30 AGATA detectors using the 30th as an external trigger. In the
30th detector a central contact energy of $1332.5\pm5$ keV was
demanded. In this manner a $\gamma$-ray multiplicity of one can be
guaranteed for the remaining 29 detectors. In figure
\ref{fig:co60gated} the $\gamma$-ray spectrum is shown, together with
spectra made with the two different treatments of the
single-interaction validation used in this work. The peak-to-total
using the empirically fitted maximum distance in germanium for single
interactions is 36.4(4)\%. It is well known that the peak-to-total in
a $\gamma$-ray tracking array is dominated by single-interaction
points accepted as events corresponding to a direct absorption of the
total $\gamma$-ray energy via the photoelectric effect. Excluding such
events the peak-to-total is increased to 52.4(6)\%, with a reduction
in efficiency for the full energy peak of 17\%.  The variation of
peak-to-total and efficiency at 1173 keV as a function of the
$P_{track}$ parameter is shown in figure \ref{fig:peaktototal}, for
the cases when single interactions are included or excluded. Note that
above $P_{track}>0.7$ no events with multiple interaction points are
left. From figure \ref{fig:peaktototal} it is clear that for the OFT
algorithm the peak-to-total has a weak dependence on the $P_{track}$
parameter, again showing that it is $\sigma_{\theta}$ that is the
most important parameter for OFT.

\begin{figure}[htbp]
  \centering
  \includegraphics [width=.5\textwidth] {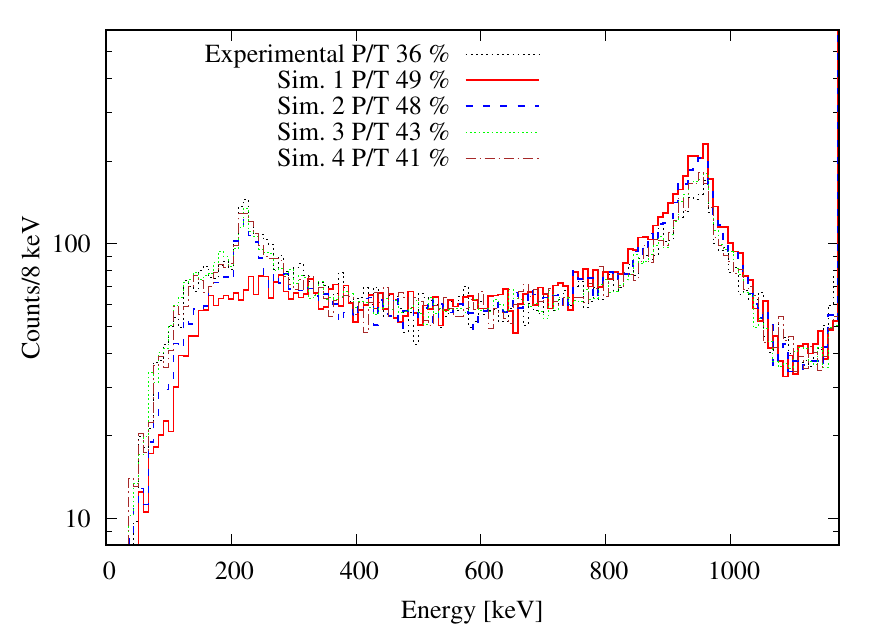}
  \caption{Comparisons between experimental spectrum (in black) and
    spectra from four different simulations. The spectra are
    normalized to the same number of counts in the region 0-1165
    keV. Simulation 1 includes only AGATA, simulation 2 also includes
    a schematic implementation of the VAMOS dipole magnet, simulation
    3 further adds concrete walls around the experimental
    setup. Finally, simulation 4 has thicker dead layers added to the
    HPGe crystals. For further details on the simulations, see the
    text. }
  \label{fig:geantsimpeaktototal}
\end{figure}

Monte Carlo simulations using the AGATA simulation package were made
in order to compare the simulated $\gamma$-ray tracking performance
with experimental data. In the simulations a $^{60}$Co was simulated
with a source strength of 5 kBq. An absolute time was used in the
simulations allowing effects such as pile-up and random coincidences
to be simulated. Gamma-ray interactions in the same segment were
packed at their energy-weighted average positions. These were then
written into the same data format as used to store experimental
post-PSA data. This allowed the use of identical $\gamma$-ray tracking
and data analyses codes for the experimental and simulated data, {\it
  i.e.}  the simulated data was treated exactly as explained for the
experimental data above.  Four different simulations were
performed. The first one including only the HPGe crystals and the
aluminum end-caps. The second simulation included a large piece of
steel to mimic the large quadrupole magnet of the VAMOS. The third
simulation included both the large piece of steel mimicking VAMOS,
concrete walls and the target chamber. A fourth simulation was also
performed adding to the third simulations thicker dead layers to the
HPGe crystals. The added dead layers were 3 mm at the back side of the
detector and 2.5 mm around the central contact. The peak-to-total for
the different simulations, when gating on the 1332.5 keV transition to
look at the 1173 keV transition were 49\%, 48\%, 43\%, and 41\%,
respectively. This is to be compared to the experimental value of
36\%.  In figure \ref{fig:geantsimpeaktototal} the Compton scattering
part of the 1332.5 keV gated $^{60}$Co spectra is shown for
experimental and simulated data. In the experimental spectrum a
pronounced back-scattering peak can be seen just above 200 keV. The
simulation labeled 1, only including AGATA itself, does not show such
a back-scattering peak and consequently the peak-to-total is much
better than for the experimental data. For simulation 2, where the
VAMOS quadrupole has been included in a very schematic way a clear
back-scattering peak emerges. However, at both lower and higher
energies as compared to the back-scattering peak the experimental data
contains more counts. In simulation 3, where the concrete walls are
included together with the scattering chamber a shape of the spectrum
very close to the experimental one is produced. This suggests that a
significant fraction of the spectrum is not due to Compton scattering
inside the HPGe crystals of AGATA, but from the scattering on the
structures around AGATA into AGATA. Including thicker dead layers in
the HPGe crystals in the simulation, as done for the fourth
simulation, increases slightly the amount of background between the
full-energy peak and the Compton edge, but does not change the shape
of the spectrum in a significant way. However, the peak-to-total is
decreased by about 5\%. These \lq\lq back scattered\rq\rq\ $\gamma$
rays are very difficult to properly discriminate against as they are
from the point of view of $\gamma$-ray tracking perfectly good single
interaction point events in the front of the crystals.

\subsection{In-beam efficiency of AGATA coupled to VAMOS}
\label{subsec:inbeameff}

The in-beam efficiency of AGATA is different from that of source
measurements, as the efficiency is also a function of count rate in
the individual detectors due to pile-up (rejected and non rejected)
and rate limitations in the electronics. In-beam efficiency varies
from experiment to experiment therefor exact numbers are both
difficult to reliably produce and not of general interest. The aim of
the section is to give a useful rule of thumb to allow consistency
checks when analysing data. The in-beam efficiency for events with a
higher $\gamma$-ray fold than one also depends on the angular
distribution of and correlation of the $\gamma$-ray transitions used
to measure it. This both via pure geometrical effects and via the
lowered $\gamma$-ray tracking efficiency for $\gamma$ rays with a
preference for being emitted in parallel.
The in-beam efficiency has been estimated for AGATA coupled to VAMOS
for an experiment where a $^{92}$Mo beam impinged on a $^{92}$Mo
target, and the beam-like reaction products were unambiguously
identified in VAMOS, also providing the velocity vector for Doppler
correction. During this experiment 23 AGATA crystals were operational
in the array, each counting at around 45 kHz with a shaping time of
2.5 $\mu$s. As the target and the beam both were $^{92}$Mo,
de-excitation of target-like and beam-like particles could be
studied. The beam-like and target-like nuclei travel with a relative
angle of about $90^{\circ}$, allowing an estimate of the effect of the
angular distribution on the measured efficiency.

\begin{figure}[htbp]
   \centering
   \includegraphics [width=.5\textwidth] {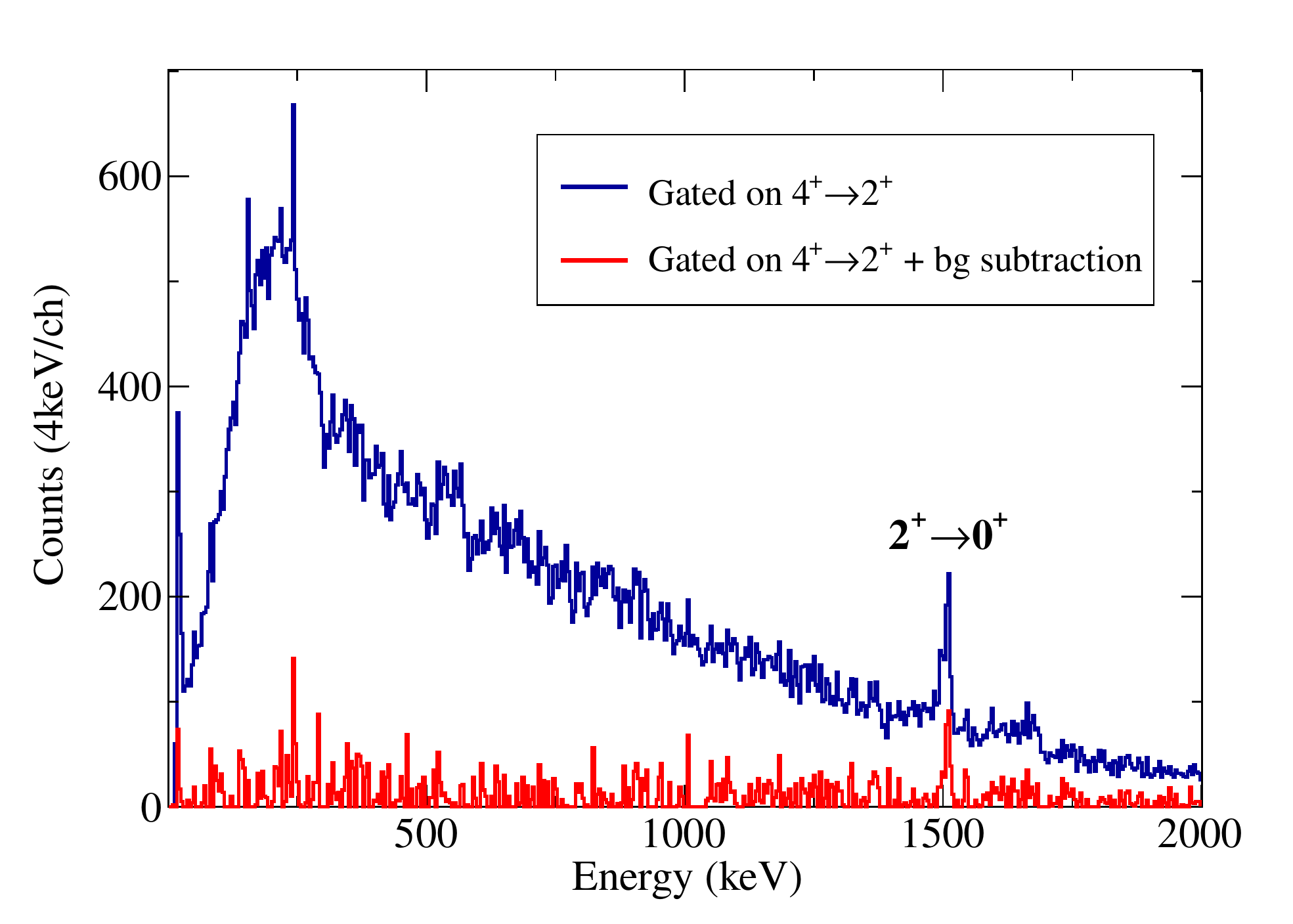}
   \caption{Gamma-ray spectra showing the 1510 keV
     $2^+_1\rightarrow0^+_1$ transition in $^{92}$Mo used to estimate
     the in beam efficiency.}
   \label{fig:ggeff}
\end{figure}

The coincidence method was used to determine the efficiency at 1510
keV,{\it i.e} , the number of detected $2^+_1\rightarrow0^+_1$
$\gamma$ rays per detected $\gamma$ ray from the
$4^+_1\rightarrow2^+_1$ transition was determined. Peak intensities
were extracted from singles spectra and from $\gamma\gamma$
coincidence matrices. The projected gate in the $\gamma\gamma$
coincidence matrix is shown in figure \ref{fig:ggeff}.
The efficiency at 1510 keV extracted using this method is after
$\gamma$-ray tracking 1.5(1)\%, to be compared with the expected
efficiency of about 2.5\% for 23 AGATA crystals at an energy of 1.5
MeV. This loss of efficiency, some 40\% lower, has several origins. In
this section we will try to identify the sources of this reduction. At
count rates of about 45 kHz and a shaping time of 2.5 $\mu$s there is
a loss of the order of 20\% due to the pile-up protection built in the
AGATA pre-processing firmware \cite{recchia2010}. There is also the
loss in tracking efficiency for higher fold events. A lower limit for
this can be estimated using the smallest used cluster angle in the OFT
of $8^{\circ}$ (as can be seen in figure \ref{AngularCorrelations} the
efficiency to track two $\gamma$ rays inside this cone is close to
zero) which corresponds to approximately 6\% of the solid angle of 23
AGATA detectors. These contributions add up to about 25\% of losses
(i.e. more than half of the lost efficiency) that are rate dependant,
via the pile up, and related to the detector physics ({\it i.e}. the
rise time of the HPGe crystals and average cluster size for typical
$\gamma$ rays) and therefore always will be present. There is an open
question from where the remaining about 15\% of efficiency loss is
coming. Measurements suggests that 5\% to 10\% could come from
overload beyond specification of the trigger distribution system
related to the high total rate (more than 1MHz).

\section{Position resolution of the PSA}
\label{sec:psares}
\begin{figure}[htbp]
   \centering
   \includegraphics [width=.5\textwidth] {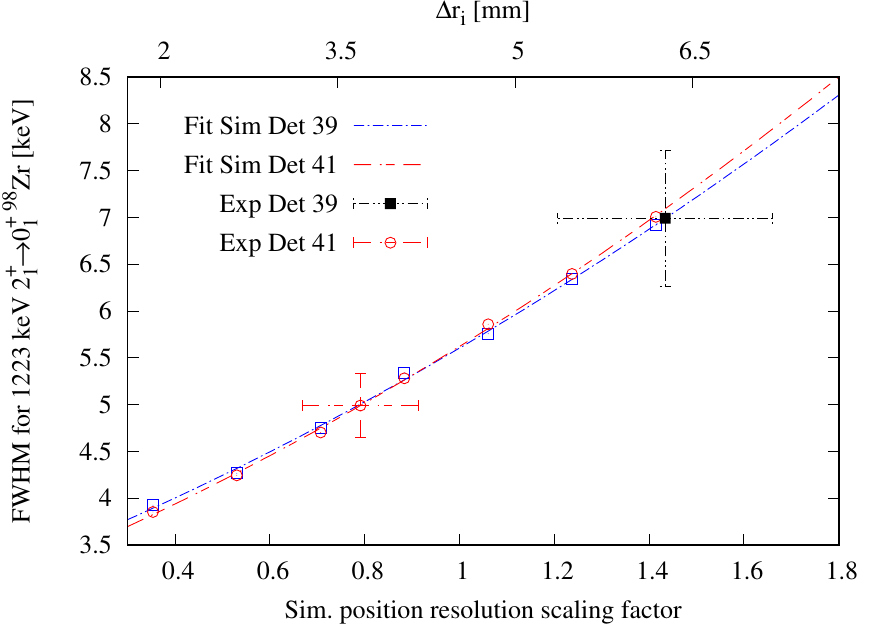}
   \caption{Estimated position resolution for two detectors in the
     AGATA array. The lines are fits to the FWHM of simulated data
     sets for the two different detectors where the assumed position
     resolution has been varied. Large symbols show the measured FWHM
     for each detector (y-axis) and corresponding deduced position
     resolution (x-axis). For details on simulation and experiment,
     see text. Note that the lower x-axis is a scaling factor of the
     position resolution given in S{\"o}derstr{\"o}m
     et. al. \cite{Soderstrom201196} for the FWHM in one
     dimension. The upper x-axis shows the average resolution for the
     interaction points used for Doppler Correction. }
   \label{fig:PSAres}
\end{figure}
The VAMOS allows for a very precise determination of the recoil vector
of the identified ion. The direction can in this context be considered
as exact whereas the velocity has an error in the order of a few per
mill. Given that the recoil velocity has a very small error the
position resolution can be estimated by the Doppler Broadening of the
$\gamma$-ray peaks via the Doppler Shift given by (for details see,
e.g., S{\"o}derstr{\"o}m et al. \cite{Soderstrom201196})
\begin{equation}
  E_{\gamma}=E_{\gamma0}\frac{\sqrt{(1-\beta^2)}}{(1-\beta cos\theta)}
\label{eq:doppler}
\end{equation}
where $E_{\gamma}$ is the energy detected in the detector,
$E_{\gamma0}$ is the energy of the $\gamma$ ray in the rest frame of
the nucleus, $\beta$ is the velocity of the nucleus emitting the
$\gamma$ ray and $\theta$ is the angle between the velocity of the
emitting nucleus and the $\gamma$ ray in the laboratory frame. From
this we have a $\gamma$-ray peak width $\Delta E_{\gamma0}$
of 
\begin{equation}
  (\Delta E_{\gamma0})^2=\left(\frac{\partial E_{\gamma0}}{\partial
        E_{\gamma}}\Delta E_{\gamma}\right)^2\
    +\left(\frac{\partial E_{\gamma0}}{\partial
        \beta}\Delta\beta\right)^2+
    \left(\frac{\partial E_{\gamma0}}{\partial
        \theta}\Delta\theta\right)^2.
\label{eq:dopplerbroadening}
\end{equation}
This can be used to evaluate the performance of the PSA via the
relation
\begin{equation}
  \cos\theta=\frac{\vec{v}\cdot\vec{r}}{|\vec{v}||\vec{r}|}
\label{eq:deftheta}
\end{equation}
where $\vec{v}$ is the recoil velocity as detected by VAMOS and
$\vec{r}$ is the position vector of the first $\gamma$-ray interaction
as given by $\gamma$-ray tracking. The method employed to determine
the position resolution for six different AGATA crystals is to perform
Geant4 simulations that in a realistic way take into account all
experimental contributions to the FWHM of the $\gamma$-ray peaks while
varying the assumed position resolution of the PSA. The experimental
FWHM of the $\gamma$-ray peak can then be used to interpolate the
actual position resolution of the PSA as done by, e.g., Recchia et
al. \cite{RECCHIA2009555}.

In this case the experiment was a fusion-fission experiment
populating, among other nuclei, $^{98}$Zr. A beam of $^{238}$U
impinged on a 10 $\mu$m think $^{9}$Be foil. The VAMOS was positioned
at 28$^{\circ}$ relative to the beam axis. Six AGATA detectors close
to $112^{\circ}$ relative to the recoil direction were used to sample
the position resolution of the detectors in the array, as they had the
largest Doppler Broadening, increasing the sensitivity to the position
resolution. As all data were analyzed after $\gamma$-ray tracking it
was the interaction used for Doppler Correction that determined which
detector was studied. The FWHM of the $\gamma$-ray peaks were
determined using Gaussian fits.  An error $\Delta\beta/\beta=0.0045$
as deduced from the mass resolution of VAMOS gives a constant
contribution to the FWHM of the $\gamma$-ray peak of 0.13\%.

The simulations took into account the energy loss in the target and
straggling as the reaction products leave the target as well as the
acceptance of VAMOS. For these simulations the AGATA geant4
simulations package was used \cite{Farnea2010331}. In the simulations
a perfect knowledge of the recoil velocity was assumed
($\Delta\vec{v}=0$). An intrinsic resolution of the AGATA detectors of
2.6 keV at 1332 keV was assumed for all detectors. Peak widths as a
function of position resolution were determined for seven different
position resolutions. As a baseline the experimentally determined
energy dependent position resolution from S{\"o}derstr{\"o}m et
al. \cite{Soderstrom201196}
\begin{equation}
    \Delta r_{i}=1.9+4.4*\sqrt{100 keV/E_{i}}\text{~mm FWHM}
\label{eq:soderstrom}
\end{equation}
where $E_i$ is the energy of the interaction point i. The resolution
was scaled with a value ranging 0.36 to 1.41 for the different
simulations. This procedure allows to correctly capture the variation
in position resolution with the energy of the interaction point used
for Doppler correction.  From the simulations it was determined that
the average position resolution for the interaction point used for the
Doppler Correction when using the non-scaled function of
S{\"o}derstr{\"o}m et al. \cite{Soderstrom201196} is 4.3 mm FWHM. For
each assumed position resolution the FWHM of the simulated
$\gamma$-ray peak for each detector was determined by a Gaussian
fit. The extra width coming from the error in recoil velocity was
added quadratically. In figure \ref{fig:PSAres} these values are shown
with small symbols for detector 39 and 41 (which has the best and
worst experimental position resolution, respectively). To each set of
FWHM coming from the variation of position resolution a second degree
polynomial function was fitted. Using the inverse of these functions
the position resolution of the individual detectors can be determined
(see large symbols in figure \ref{fig:PSAres}). Note that in figure
\ref{fig:PSAres} the x-axis is a scaling factor with the previously
determined position resolution as base, {\it i.e.} 1 means the
detector has the same PSA performance that was previously
measured. The six detectors used to sample the position resolution are
located in the span 0.79-1.4 (as compared to S{\"o}derstr{\"o}m et
al. \cite{Soderstrom201196}), with five detectors larger than 1.08 and
a weighted average of 1.15. This corresponds to an average position
resolutions used for Doppler Correction of 3.7 mm-6.1 mm FWHM, with a
weighted average of 5.1 mm FWHM. The average error on the estimated
position resolution is 1 mm. There is no obvious difference in how the
detectors perform for other parameters than the PSA, nor in how they
have been treated. It should be noted that the probability of having a
maximum difference of 2.4 between six values randomly taken from a
Gaussian distribution with a $\sigma=1$ is in the order of 50$\%$,
i.e. our results is rather probable even if all the detectors are
performing identically. It is however of interest in a future work to
investigate the variance of detector performance with respect to PSA
in AGATA.

\section{Angular Correlations in AGATA}
\label{sec:angcorr}

The use of AGATA for angular correlation measurements to determine the
multipolarity of $\gamma$ decays has been investigated using source
data. Two pairs of $\gamma$-$\gamma$ cascades from the decay of
\(^{152}\)Eu were used: The first pair was the 1408 keV-121.8 keV
coincidence in \(^{152}\)Sm de-exciting the
2\(^{\text{-}}_{\text{1}}\) level at 1530 keV to to the ground state
via the 2\(^{\text{+}}_{\text{1}}\) level at 121.8 keV. The second
pair is the 244.7 keV-121.8 keV de-exciting the
4\(^{\text{+}}_{\text{1}}\) level at 366.5 keV and the
2\(^{\text{+}}_{\text{1}}\) level, also in \(^{152}\)Sm.

\begin{figure}[htbp]
\centering
\includegraphics[width=.5\textwidth]{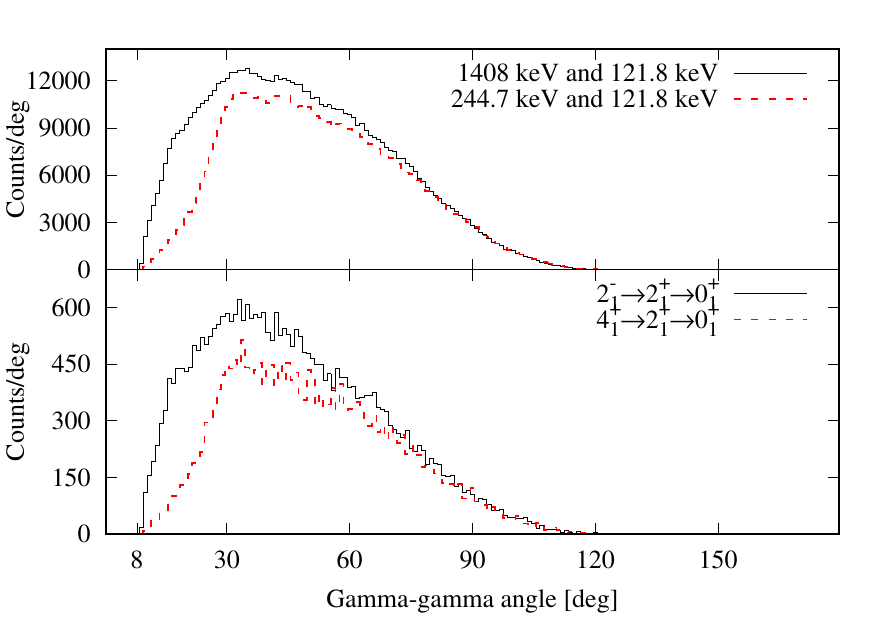} 
\caption{\label{fig:org7c0be69} Histograms used for angular
  correlation measurements using AGATA. The lower panel shows the
  angle between the two correlated $\gamma$ rays detected in
  AGATA. The upper panel shows the angle between $\gamma$ rays from
  uncorrelated events concatenated before tracking.}
\end{figure}

The tracking algorithm identifies the first interaction point of each
\(\gamma\) ray and as the position of the source is known the angle
between the \(\gamma\) rays in the 1408 keV-121.8 keV pair and the
244.7 keV-121.8 keV pair could be determined and histogramed, see
lower panel in figure \ref{fig:org7c0be69}. The main features of the
two pairs of \(\gamma\) rays are similar. The cut at about 8 degrees
is a result of the tracking algorithm, whereas for larger angles the
geometry of AGATA as used for the source measurement dominates the
shape of the spectra. The slower rise in intensity for the 244.7-121.8
keV cascade at low angles comes from the intrinsic difficulty to track
two low-energy $\gamma$ rays emitted into a small solid angle, since
they often will be reconstructed as one $\gamma$ ray with sum
energy. The angular correlation is then extracted by normalizing for
geometrical effects and the already mentioned decrease in efficiency
for two low-energy $\gamma$ rays absorbed close to each other. The
normalisation was created by tracking events consisting of the
interaction points of two events each with a total energy
corresponding to one of the $\gamma$ rays in the cascade of interest
concatenated into one event, thus generating pairs of $\gamma$ rays
with the correct energies, but with no angular correlation. From the
tracked events the angle between the $\gamma$ rays was then again
extracted. The resulting histograms for the two pairs of $\gamma$ rays
are shown in the upper panel of figure \ref{fig:org7c0be69}.

\begin{figure}[htbp]
\centering
\includegraphics[width=.5\textwidth]{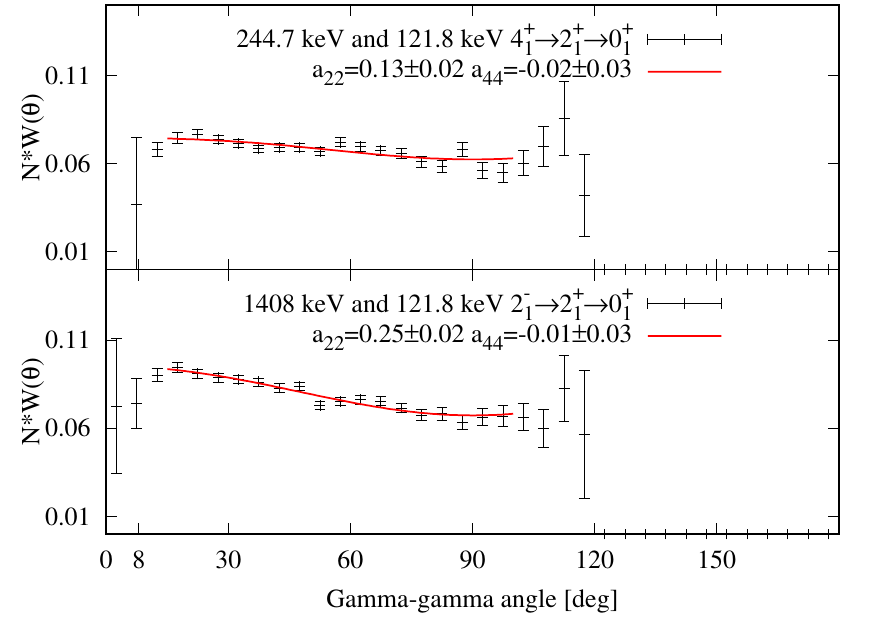}
\caption{\label{fig:orgd0fbfd3} Gamma-gamma angular correlations
  measured with AGATA. The upper panel shows the angular correlation
  for the
  4\(^{\text{+}}_{\text{1}}
  \rightarrow\)2\(^{\text{+}}_{\text{1}}
  \rightarrow\)0\(^{\text{+}}_{\text{1}}\) pair of transitions in
  $^{152}$Sm. The lower panel is for the
  2\(^{\text{-}}_{\text{1}}
  \rightarrow\)2\(^{\text{+}}_{\text{1}}
  \rightarrow\)0\(^{\text{+}}_{\text{1}}\) transitions}
\end{figure}

By dividing the histograms in the lower panel in figure
\ref{fig:org7c0be69} by the upper panel the histograms shown in figure
\ref{fig:orgd0fbfd3} are created. The upper panel is for the
4\(^{\text{+}}_{\text{1}}
\rightarrow\)2\(^{\text{+}}_{\text{1}}
\rightarrow\)0\(^{\text{+}}_{\text{1}}\) cascade, the lower panel for
the 2\(^{\text{-}}_{\text{1}}
\rightarrow\)2\(^{\text{+}}_{\text{1}}
\rightarrow\)0\(^{\text{+}}_{\text{1}}\) cascade. For each angular
correlation the expression
\begin{equation}
  W(\theta)=1+a_{22}P_2(cos(\theta))+a_{44}P_4(cos(\theta))
\label{eq:angcorr}  
\end{equation}
where $a_{22,44}$ are the directional correlations coefficients and
$P_{2/4}$ are the Legendre polynomials of order 2 and 4 respectively,
have been fitted, and the $a_{22}$ and $a_{44}$ coefficients
extracted. For the stretched
4\(^{\text{+}}_{\text{1}}
\rightarrow\)2\(^{\text{+}}_{\text{1}}
\rightarrow\)0\(^{\text{+}}_{\text{1}}\) cascade the fitted values are
$a_{22}=0.13\pm0.02$ and $a_{44}=-0.02\pm0.03$ to be compared with
theoretical values of $a_{22}=0.102$ and $a_{44}=0.0091$. For the
non-mixed the
2\(^{\text{-}}_{\text{1}}
\rightarrow\)2\(^{\text{+}}_{\text{1}}
\rightarrow\)0\(^{\text{+}}_{\text{1}}\) cascade our fit gives
$a_{22}=0.25\pm0.02$ and $a_{44}=-0.01\pm0.03$, for which the
theoretical values are $a_{22}=0.25$ and $a_{44}=0$. With three out of
four values within $1\sigma$ this is in agreement with what is
expected if AGATA is correctly reproducing the angular correlations.

\begin{figure}[htbp]
\centering
\includegraphics[width=.5\textwidth]{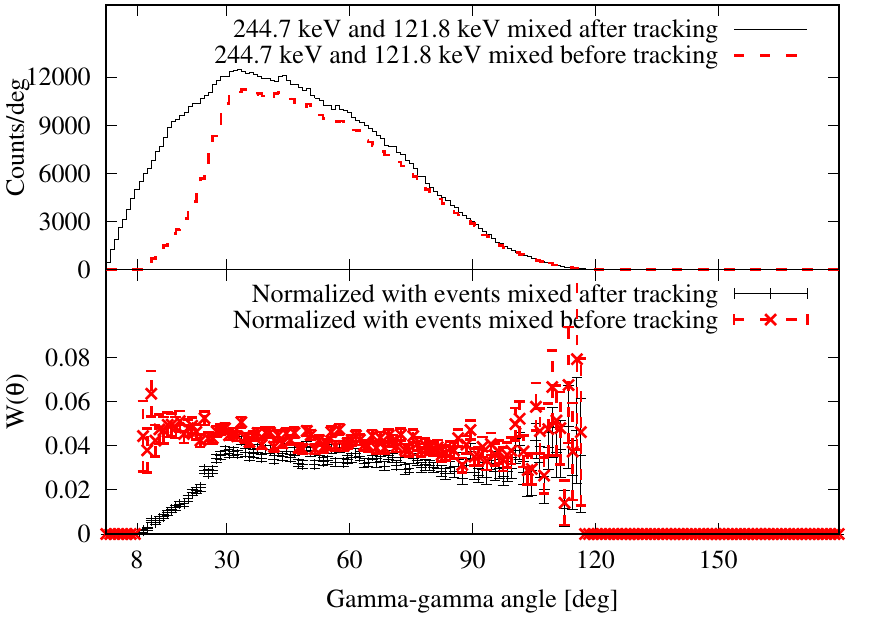}
\caption{\label{fig:orgfaa9a5c} Normalisation histograms and angular
  correlations for
  4\(^{\text{+}}_{\text{1}}
  \rightarrow\)2\(^{\text{+}}_{\text{1}}
  \rightarrow\)0\(^{\text{+}}_{\text{1}}\) using the "concatenate
  before tracking" and "concatenate after tracking" methods.}
\end{figure}

A particularity of a $\gamma$-tracking array as compared to a
classical multi-detector $\gamma$-ray spectrometer is the continuous
variation in efficiency with the angle between the detected $\gamma$
rays. For angular correlations this means the normalisation of the
angular correlations need not only to consider geometrical coverage.
This can be seen by looking at the difference between using
uncorrelated hits that are concatenated and tracked or tracked
uncorrelated $\gamma$ rays concatenated into events when constructing
the normalisation used to extract the angular correlations from the
experimental correlations. In the top panel of figure
\ref{fig:orgfaa9a5c} the histogram drawn with a black solid line shows
the distribution of $\theta$ angles between uncorrelated $\gamma$ rays
concatenated after tracking. The histogram drawn with red dashed line
shows the $\theta$ angle distribution if one instead concatenates
uncorrelated events using the individual hits and then preforms the
tracking.
The bottom panel shows the resulting angular correlations using the
two different methods of generating the normalisation. It is clear
that the effects of $\gamma$-ray tracking, included when events are
concatenated before tracking, are needed for a correct normalisation
over the entire angle range. This procedure works well for source data
where the peak-to-background ratio is high and $\gamma$ rays from
different decays have no correlations. The application of this method
to in-beam data is however more problematic due too the lower
peak-to-background ratio and cross-event angular correlations coming
from aligned nuclei.

\section{Conclusions and perspective}
\label{sec:conclusions}

The performance of AGATA installed at GANIL, coupled to the VAMOS has
been characterized. The efficiency of AGATA, as a whole as well as for
individual crystals, has been determined using both singles
measurements and coincidence methods. It has been done both using
AGATA as a standard array and as a $\gamma$-ray tracking array.  A
total efficiency for AGATA of 3.8(1)\% at 1332 keV for the nominal
geometry when using $\gamma$-ray tracking was determined. This is to
be compared to 2.9\% at 1332 keV if AGATA is used as normal
multi-detector array.  It is also shown how the efficiency extracted
from coincidence has to be corrected for angular correlation effects,
as the increased probability to emit parallel $\gamma$ rays combined
with the clustering stage of $\gamma$-ray tracking generates a loss of
efficiency that depends on the angle between the two $\gamma$
rays. This correction has to be made on top of the typical correction
made for angular correlations effects.

As the AGATA detectors have been used in different campaigns the
segments of some of the detectors are showing clear signs of neutron
damages. Annealing procedures are complicated for these detectors and
ingenious neutron damage correction procedures have been developed
allowing an almost full recovery of the intrinsic energy
resolution. The average central contact energy resolution for AGATA
(beginning of 2016) is 2.57 keV at 1332 keV and for the segments 3.08
keV at 1332 keV. Corresponding values before applying neutron damage
correction are 3.08 keV and 5.22 keV for central contacts and
segments, respectively. The neutron damage correction procedure is
effective, but at some point the detectors will need to be
annealed. Maintenance, such as annealing, are as important for the
future of AGATA as the more appealing technical developments that can
be made.

The position resolution given by the PSA for AGATA at GANIL has been
estimated for 6 AGATA crystals using data from an experiment performed
in the first half of the campaign. This was done by comparing
experimental data with Monte Carlo simulations in which the position
resolution was varied. It turns out that the average position
resolution found was a factor of 1.16(5) larger than what was measured
in a dedicated experiment \cite{Soderstrom201196}.

As the number of crystals in AGATA increases the interest in using
AGATA for angular correlations and distributions increases. Using a
$^{152}$Eu source angular correlations have been produced and methods
to properly normalize for the combined effect of geometry and
$\gamma$-ray tracking have been devised. 

Finally, the AGATA detector system is performing very well, as proven
by the physics results that have been produced. However, improvements
in the PSA and further tuning of $\gamma$-ray tracking algorithms
would be beneficial. A better understanding of the details of the
signal generation in the segmented detectors is needed to improve the
PSA. This would also allow for better handling of multiple
interactions in one segment and removing the nonphysical clustering of
interaction points. Such improvements would allow for an increased
peak-to-total.

\section{Acknowledgments}

The authors would like to thank the AGATA collaboration and the GANIL
technical staff. Gilbert Duch{\^e}ne is thanked for providing the
in-beam data set used to extract the position resolution of the
pulse-shape analysis. The excellent performance of the AGATA detectors
is assured by the AGATA Detector Working group.
This work was partially supported by the Ministry of Science,
Spain, under the Grants BES-2012-061407, SEV-2014-0398,
FPA2017-84756-C4 and by the EU FEDER funds.
The research and development on AGATA was supported by the German BMBF
under Grants 06K-167, 06KY205I, 05P12PKFNE, 05P15PKFN9, and
05P18PKFN9.
The AGATA project is supported in France by the CNRS and the CEA. 
The UK Science and Technology Facility Council (STFC) supports the
AGATA project. 

\appendix
\section{Detector and crystal positions and ids}
\label{app:detid}

\begin{table}[htp]
\begin{center}
  \caption{Crystal lookup table for the 32 crystals present in the
    set-up, although only 30 were used in the measurement. The
    capsules in position 11A and 14A were not operational and shown in
    italic. Neutron damage correction was performed on detectors
    marked in bold.}
\label{table:DetId}
\begin{tabular}{|c|c|c|c|c|}
\hline
Cluster & Crystal A & Crystal B & Crystal C & Position Array \\
\hline
ATC6&{\bf A001}&{\bf B004}&{\bf C010}&00\\
\hline
ATC8&A009&B005&{\bf C008}&02\\
\hline
ATC5&A005&{\bf B002}&{\bf C009}&03\\
\hline
ATC9&A004&B008&C013&04\\
\hline
ATC10&A010&B012&C012&05\\
\hline
ADC9&-&{\bf B011}&C011&09\\
\hline
ATC2&{\bf A003}&{\bf B003}&{\bf C005}&10\\
\hline
ATC7&\it{A006}&{\bf B013}&{\bf C006}&11\\
\hline
ATC3&{\bf A002}&{\bf B010}&{\bf C001}&12\\
\hline
ATC4&{\bf A007}&{\bf B007}&{\bf C007}&13\\
\hline
ATC1&\it{A008}&{\bf B001}&{\bf C003}&14\\
\hline             
\end{tabular}
\end{center}
\end{table}%

\end{document}